\documentclass[journal=jctcce,manuscript=article,hyperref,layout=twocolumn]{achemso}
\setkeys{acs}{doi=true}
\setkeys{acs}{chaptertitle=true}
\usepackage[version=3]{mhchem} 
\usepackage[table,xcdraw]{xcolor}
\usepackage{comment}
\usepackage{threeparttable}
\usepackage{booktabs}
\usepackage{multirow}
\usepackage{amsmath}
\usepackage{orcidlink}
\usepackage{graphicx}
\makeatletter
\newcommand*\doi{\begingroup\let\do\@makeother\dospecials
  \catcode`\{=1 \catcode`\}=2 \doi@aux}
\newcommand*\doi@aux[1]{\endgroup\href{https://doi.org/#1}{\nolinkurl{#1}}}
\makeatother

\newcommand{\kcalmol}{\mbox{kcal$\cdot$mol$^{-1}$}}

\author{Vladimir Fishman\orcidlink{0009-0004-7570-136X}}
\author{Jan M.L. Martin\orcidlink{0000-0002-0005-5074}}
\affiliation{Department of Molecular Chemistry and Materials Science, Weizmann Institute of Science, 7610001 Re\d{h}ovot, Israel}    

\email{gershom@weizmann.ac.il}
\phone{+972 8 9342533}
\fax{+972 8 9343029}

\date{\today}

\title[pANO-F12]{pANO-F12: An atomic natural orbital-inspired route to more compact basis sets for F12 explicitly correlated methods}
\abbreviations{pANO-F12}

\begin{document}
\maketitle

\begin{center}
\small \today
\end{center}

\begin{abstract}
Explicitly correlated methods such as MP2-F12 and CCSD(F12*) exhibit much faster basis set convergence (asymptotically $\propto L^{-7}$, with $L$ the highest angular momentum) than orbital-only approaches. Yet it has been pointed out that cc-pV$n$Z-F12 basis sets themselves are substantially larger than the corresponding cc-pV$n$Z, and specifically that cc-pVDZ-F12 is the size of cc-pVTZ. 
Atomic natural orbital (ANO) basis sets [J. Almlöf and P. R. Taylor, JCP 86, 4070 (1987)] constitute one way to generate compact basis sets in an orbital-only context. However, obtaining the required first-order reduced density matrix while properly accounting for the F12 geminal is problematic. In this work, we 
show that an energy minimization-based contraction process under linear independence constraints yields `pseudo-ANO' (pANO) basis sets that are functionally equivalent in quality. Subsequently, we apply this recipe to obtain pANO-F12 basis sets for H, B--F, and Al--Cl, then validate them for several benchmarks (thermochemistry, noncovalent interactions, harmonic frequencies) and for the hypersensitive out-of-plane vibrations of benzene.
We show that, unlike cc-pV$n$Z-F12,
pANO-F12 offers a route to more compact F12 basis sets more amenable to medium-sized systems, especially in conjunction with localized pair natural orbital approaches.
Overall, the pANO approach is most beneficial for the smaller double- and triple-zeta basis sets, offering either superior performance to cc-pV$n$Z-F12 at the same cost, or comparable performance at lower cost.
\end{abstract}

\section{Introduction}

Almost a century ago, Hylleraas\cite{Hylleraas1929} first proposed using an explicit interelectronic distance $r_{12}$ as a basis function for the ground state energy of helium. A few decades later, Pekeris leveraged WEIZAC (the first-ever computer in the Middle East) to obtain accurate ground-state energies of helium and helium-like atomic ions.\cite{Pekeris1958,Pekeris1959,Pekeris1962Hminus,Pekeris1962ExcitedS}

S. Francis Boys is best known as the first to introduce Gaussian basis functions in electronic structure theory.\cite{Boys1950} Later, together with his then-student Nicholas C. Handy, he proposed the so-called `transcorrelated' method\cite{BoysHandy1969FullCorr,BoysHandy1969Indeterminacy,BoysHandy1969Neon,BoysHandy1969LiH}, in which the Schr\"odinger equation is similarity-transformed  using a Jastrow factor\cite{Jastrow1955}. (See Ref. \citenum{Handy1972TCGaussians} for an early review.) After lying dormant for decades, transcorrelated approaches have started experiencing a modest revival. (See, e.g., Alavi, Kats and coworkers\cite{Alavi2021} and references therein.)

Modern explicitly correlated quantum chemistry got its start with the classic papers of Klopper and Kutzelnigg\cite{KutzelniggKlopper1991R12I,TermathKlopperKutzelnigg1991MP2R12} and Noga and Kutzelnigg\cite{NogaKutzelnigg1994CCR12}. 
In a landmark paper, Kutzelnigg and Morgan\cite{KutzelniggMorgan1992Rates} (see also earlier work by Kutzelnigg\cite{Kutzelnigg1985R12}) showed that in the presence of $r_{12}$-type `geminal' terms, basis set convergence for Gaussian basis functions in terms of the highest angular momentum $L$ is asymptotically $\propto L^{-7}$, compared to $\propto L^{-3}$ in a pure orbital basis set. (The term `geminal', from \textit{Gemini},  appears to have been first introduced for an electron pair function by Shull\cite{Shull1959firstUseOfGeminal} before Kutzelnigg\cite{Kutzelnigg1985R12} gave it the specific modern meaning of an $r_{12}$-dependent basis function.)

The bottleneck caused by the need to evaluate four-electron integrals was resolved by Klopper and Samson\cite{klopper2002explicitly} through the introduction of an auxiliary basis set. Later, Valeev\cite{Valeev2004CABS} introduced the CABS (complementary auxiliary basis set) as a way of enhancing R12 and F12 performance. (CABS also permits a one-shot correction to the SCF reference energy:\cite{Valeev2004CABS} without it, one has the topsy-turvy situation in which the SCF energy converges more \emph{slowly} with the basis set than the R12 --- or F12, \emph{vide infra} --- correlation energy!)

The transition from R12 to the present \emph{de facto} standard of F12 began in 2004, when Ten-No\cite{TenNo2004STG} first proposed Slater-type geminals (STG). In computational practice, codes such as MOLPRO,\cite{Molpro2020} TURBOMOLE,\cite{Turbomole2020,Turbomole2023} or MRCC\cite{MRCC2020,MRCC2025} approximate the STG as a linear combination of (usually six) Gaussian-type geminals (not unlike the approximation of a Slater-type orbital as a linear combination of Gaussians) as advocated a decade earlier by Persson and Taylor\cite{PerssonTaylor1996GTG}. 

Two very detailed reviews of explicitly correlated methods can be found, back to back, in H\"attig et al.\cite{HaettigKlopperKoehnTew2012ChemRev} and Kong et al.\cite{KongBischoffValeev2012ChemRev}; for a somewhat more recent perspective, see Gr\"uneis et al.\cite{Grueneis2017Perspective}.

More recently, F12 methods have been paired with localized natural orbital approaches, thus potentially offering a `best of both worlds' accurate wavefunction approach for large systems. (See Refs. \citenum{MaWerner2018PNOF12Review,Tew2021PrincipalDomainsF12} for reviews.) Mehta and Martin have also shown\cite{jmlm314,jmlm318} that F12 is highly beneficial for evaluating the MP2-like term in double-hybrid density functional methods (see Ref. \citenum{jmlm290} for a review). 

It became clear early on\cite{peterson2008systematically} that existing Gaussian basis sets for accurate correlated \emph{orbital} calculations --- such as the Dunning `correlation consistent' family,\cite{dunning1989gaussian,Peterson2011EIBC_eibc0408,Jensen2013AtomicOrbitalBasisSets,Hill2013GaussianBasisSetsMolecularApplications,NagyJensen2017BasisSetsRiCC} which has become something of an industry standard for wavefunction \emph{ab initio} theory --- are rather suboptimal for F12 calculations. This is easily understood thus: any basis set optimized for orbital-only correlation will be skewed towards the energetically important short-range correlation, which in an F12 calculation is already covered by the geminals. 

Peterson et al.\cite{peterson2008systematically} first presented the cc-pV$n$Z-F12 basis sets (correlation consistent polarized valence $n$-tuple zeta, $n$=D,T,Q) and demonstrated their superior performance. Later, Sylvetsky et al.\cite{jmlm269} showed that, in computational thermochemistry applications of F12 methods, dedicated cc-pV$n$Z-F12 basis sets  greatly reduce basis set superposition error (BSSE) and hence do not suffer from the nonmonotonic basis set convergence observed\cite{jmlm269} for the ordinary aug-cc-pV$n$Z basis sets. Later, cc-pV$n$Z-F12 and cc-pV$n$Z-F12-PP (pseudopotential\cite{DolgCao2012RelativisticPseudopotentials}-based) versions for much of the periodic table have been published,\cite{Hill2014,Hill2021,jmlm325} as well as cc-pV5Z-F12 for high-accuracy calculations.\cite{jmlm261,jmlm269}

A commonly voiced objection by applied quantum chemists against F12 methods comes down to `F12 basis sets are so much larger --- cc-pVDZ-F12 is the size of cc-pVTZ, so what does one gain?' or some paraphrase of this argument. This prompts the question: are cc-pV$n$Z-F12 truly the most \emph{compact}  basis sets possible for F12 calculations? (We define `most compact' here as: maximum correlation energy recovery for minimal size.)

For orbital-based calculations, Alml\"of and Taylor\cite{almlof1987general} presented the so-called atomic natural orbital (ANO) basis sets. Briefly summarizing:
\begin{itemize}
    \item take a large primitive Gaussian basis set that spans an adequately broad range of exponents in each angular momentum;
    \item carry out an atomic configuration interaction calculation in that primitive set;
    \item obtain the 1PDM (one-particle reduced density matrix) and diagonalize it: the eigenvectors are the natural orbitals (NOs), and the corresponding eigenvalues are a generalization of the occupation number to correlated wave functions;
    \item sort the NOs by descending occupation number.
\end{itemize}
They then observed that the occupation numbers form clusters, naturally giving rise to contraction patterns such as ANO321, ANO4321, and ANO54321, where the digits $m$ refer to the first $m$ contractions of $s$, $p$, $d$,... symmetry, respectively. Note also that no tedious exponent optimization steps are involved.

With sufficiently large primitive sets, ANOs of a given contraction pattern will be the ones that minimize the BSSE for that contraction. There are many indications (e.g., Lee and Taylor\cite{jmlm099,jmlm104} in the late 1990s; McCaslin and Stanton\cite{McCaslin2013}; Neese and Valeev\cite{NeeseValeev2010})  that ANOs are much more resilient to BSSE than cc basis sets of equivalent size.

This raises the tantalizing prospect that a form of ANO-F12 basis sets might be more compact than cc-pV$n$Z-F12. However, obtaining proper 1PDMs in the presence of F12 is at present not a practical option.

But can the need for an 1PDM be circumvented? In the present work, we will first show \emph{in an orbital context} that energy minimization of contraction coefficients will, with judicious constraints, yield pseudo-ANO (pANO) basis sets functionally equivalent to proper ANOs of the same size.

Having thus established an energy optimization-based alternative for ANOs, we will then apply the pANO concept to the optimization of F12 basis sets, yielding pANO-F12.

We then finally put the obtained pANO-F12 basis sets to the test for practical thermochemistry, noncovalent interactions, and vibrational frequencies. We will show that especially for smaller sets, pANO-F12 are significantly more compact for equivalent performance.

As a by-product, we also propose diffuse function-augmented aug-pANO-F12 basis set for application to anionic systems. We will also investigate whether, for neutral systems, they can further reduce BSSE and its concomitant artifacts.

\section{Computational Details}
\label{sec:methods}

Most of the calculations were carried out with MOLPRO\cite{Molpro2020} version 2026.1; some of the initial work was carried out using the earlier version 2024.1. Integral screening thresholds and convergence criteria were set to essentially machine precision, 10$^{-12}$ $E_h$ for the energy, 10$^{-28}$ for two-electron integral screening, and 10$^{-30}$ for shell pre-factor screening. 

Conventional natural orbitals were obtained by diagonalizing the 1PDM 
of an atomic CISD calculation from a CSF (configuration state function) reference. For all first and second row elements except nitrogen and phosphorus (which have single-determinant $^4S$ ground states), the SCF calculations for the $^2P$ ground states of \{B, F, Al, Cl \} and the $^3P$ ground states of \{C, O, Si, S\} were carried out for a 3-determinant CSF (configuration state function)  to ensure spherical symmetry, and hence a first-order reduced density matrix that is properly blocked by angular momenta. For instance, for carbon atom, the $^3P$ ground state is represented as a linear combination of the three symmetry-equivalent determinants. $(|\uparrow|\uparrow|0>+|\uparrow|0|\uparrow>+|0|\uparrow|\uparrow>)/\sqrt{3}$.  For the nitrogen and phosphorus $^4S$ ground states, with their half-filled $p$ shells, a single $|\uparrow|\uparrow|\uparrow>$ determinant suffices. 

Some calculations were carried out using the rigorous CCSD(F12*) coupled cluster method\cite{Kohn2010CCSDF12star} in MOLPRO version 2026.1 (available there under the name \texttt{CCSD-F12c}); triple excitations were subjected to Marchetti-Werner scaling\cite{Marchetti2009} as indicated by the CCSD(T*)(F12*) label. Some larger calculations were carried out using the PNO-LCCSD-F12b\cite{MaWerner2018PNOF12Review} implementation in MOLPRO. 

All calculations were run on CHEMFARM, which is the  HPC facility of the Faculty of Chemistry at the Weizmann Institute of Science.

   Unless indicated otherwise, the auxiliary basis sets were as follows. For cc-pV$n$Z-F12 and pANO-$n$Z-F12, the cc-pV$n$Z-F12 auxiliary sets were used (cc-pV$n$Z-F12/OptRI\cite{YousafPeterson2008OptRI} as the RI basis for constructing the CABS, and the corresponding MP2FIT and JKFIT sets for density fitting of MP2-like and Coulomb-exchange integrals, respectively); for aug-pANO-$n$Z-F12, the corresponding aug-cc-pV$n$Z-F12 auxiliary sets in the MOLPRO library. For conventional aug-cc-pV$n$Z and aug-cc-pV($n$+d)Z basis sets, the MOLPRO defaults were used, which (perhaps surprisingly, but possibly for backward compatibility reasons) employ aug-cc-pV$n$Z/JKFIT rather than aug-cc-pV$n$Z/OptRI\cite{YousafPeterson2009augOptRI} as the RI basis.
   

\subsection{\label{sec:opt-contracttion-coeff}Optimization of contraction coefficients}

Once the natural orbitals were obtained, we proceeded trying to generate their pseudo-ANO (pANO) counterparts. For the first pANO of any symmetry, we set to unity the coefficient for the primitive that we expect to be most prominent, and obtain the remaining contraction coefficients by energy minimization using the Nelder-Mead SIMPLEX derivative-free method\cite{nelder1965simplex}. If one of the resulting optimized coefficients was larger than one, we fixed just \emph{that} primitive's coefficient to 1 instead and repeated the optimization.

For the next contracted function, the coefficient that had been frozen to 1 in the previous step was instead frozen to 0, thus ensuring linear independence (albeit not orthogonality). The process of adding a contraction was repeated until we ran out of primitives with nontrivial coefficients.

\subsection{A remark on JK-fit basis sets for second-row elements}

In the course of this study, we found unexpected discrepancies (up to ca. 0.3 kcal/mol) between the  MP2-F12/V5Z-F12 total atomization energies of second-row molecules in the present work and those previously reported in the ESI of Mehta and Martin\cite{jmlm312}. Further investigation revealed that similar discrepancies also existed between RI-MP2-F12/V5Z-F12 and MP2-F12/V5Z-F12, and that the SCF energy accounted for most of them.
(For the sensitivity of the SCF component of F12 calculations to the auxiliary basis sets, see also Shaw and Hill.\cite{ShawHill2017CABSsingles})

Replacing the aug-cc-pV5Z-JKFIT Coulomb--exchange fitting basis set\cite{Weigend2002_RIHF_direct,Weigend2006_coulomb_fitting} by a gargantuan `reference-JK' set taken from Ref. \citenum{Hill2009_extrap_MP2_CCSD_F12} largely suppressed the discrepancy. As it turned out, the deficiency in standard aug-cc-pV5Z-JKFIT could be remedied much more economically by substituting aug-cc-pV(5+d)Z-JKFIT,\cite{Nash2023_ccpVndZ_JKFit}  which has one tighter primitive each of $f$ and $g$ symmetries.

\section{Results and Discussion}

\subsection{Proof of Concept: equivalence of ANO and pANO}

First, we ought to confirm that pANOs are in fact equivalent in quality to their ANO counterparts. We can do so within an orbital-only framework. In fact, we can carry out a three-way comparison: pANO vs. ANOs vs. the equivalent number of \textit{uncontracted} Gaussians.

We started with the primitives of the very large 7ZaPa set of Ranasinghe and Petersson\cite{ranasinghe2013ccsd}, from which we truncated the $k$ functions (as MOLPRO's integral code has a hard upper limit of $i$ functions). This leaves us with a $(25s,20p,7d,6f,5g,4h,3i)$ primitive set for first-row elements, and $(29s,25p,8d,6f,5g,4h,3i)$ for second-row elements. In the original basis set, these are contracted to  $[9s,8p,7d,6f,5g,4h,3i]$ and $[10s,9p,8d,6f,5g,4h,3i]$, respectively --- where it should be noted that the $[\{7,8\}d,6f,5g,4h,3i]$ `polarization' (really, angular correlation) functions were actually just the uncontracted primitives. 

CISD with symmetry-adapted CSF references for the ground-state atoms was carried out for these basis sets, and the NOs obtained by diagonalization of the 1PDM used as contraction coefficients. Energies were then calculated incrementally, adding successive contractions (ANO or pANO) one at a time.
In order not to `clutter' the comparison needlessly, we used the same large uncontracted $sp$ set throughout and focused on contracting the $d, f, g, h, i$ angular momenta. 

Tables \ref{tab:pANO-ANO-diffs-row1} and \ref{tab:pANO-ANO-diffs-row2} demonstrate that for each polarization function the energy differences between corresponding ANO and pANO contractions start off small for the first 1-2 contractions and taper off into nothingness as contractions are successively added. Total energies for pANO are typically slightly lower than for the equivalent ANO contraction, the few exceptions being in the microhartree range.

\begin{table*}[!h]
\caption{$E$(pANO)$-E$(ANO) (in hartree) for the first-row elements.}
\label{tab:pANO-ANO-diffs-row1}
\begin{tabular}{|l|l|l|l|l|l|}
\hline\hline
 &  \multicolumn{5}{c|}{Differences} \\
\hline
Contraction size & Boron & Carbon & Nitrogen & Oxygen & Fluorine \\
\hline
1d & -0.000010 & -0.000031 & -0.000089 & -0.000362 & -0.000853 \\
2d & -0.000013 & -0.000026 & -0.000042 & -0.000128 & -0.000225 \\
3d & -0.000002 & -0.000008 & -0.000017 & -0.000036 & -0.000060 \\
4d & -0.000002 & -0.000003 & -0.000003 & -0.000017 & -0.000035 \\
5d &  0.000000  & 0.000000 & 0.000000 & 0.000005  & 0.000006  \\
6d &  0.000000  & 0.000000 & 0.000000 & 0.000000  & -0.000003 \\
\hline
1f & -0.000011 & -0.000028 & -0.000060 & -0.000188 & -0.000335 \\
2f & -0.000005 & -0.000013 & -0.000027 & -0.000079 & -0.000139 \\
3f & -0.000001 & -0.000004 & -0.000010 & -0.000025 & -0.000045 \\
4f & -0.000001 & -0.000002 & -0.000002 & -0.000010 & -0.000038 \\
5f & 0.000000  & 0.000000  & 0.000000  & 0.000001  & 0.000001  \\
\hline
1g & -0.000006 & -0.000016 & -0.000034 & -0.000085 & -0.000146 \\
2g & -0.000003 & -0.000008 & -0.000017 & -0.000046 & -0.000084 \\
3g & -0.000001 & -0.000003 & -0.000005 & -0.000017 & -0.000023 \\
4g & 0.000000  & -0.000003 & -0.000005 & -0.000029 & -0.000041 \\
\hline
1h & -0.000003 & -0.000009 & -0.000018 & -0.000045 & -0.000078 \\
2h & -0.000002 & -0.000005 & -0.000010 & -0.000026 & -0.000052 \\
3h & -0.000001 & -0.000002 & -0.000004 & -0.000012 & -0.000012 \\
\hline
1i & -0.000002 & -0.000005 & -0.000010 & -0.000025 & -0.000041 \\
2i & -0.000001 & -0.000002 & -0.000005 & -0.000013 & -0.000034 \\
\hline\hline
\end{tabular}
\end{table*}

\begin{table*}[!h]
\caption{$E$(pANO)$-E$(ANO) (in hartree) for the second-row elements.}
\label{tab:pANO-ANO-diffs-row2}
\begin{tabular}{|l|l|l|l|l|l|}
\hline\hline
 &  \multicolumn{5}{c|}{Differences} \\
\hline
Contraction size & Aluminium & Silicon & Phosphorus & Sulfur & Chlorine \\
\hline
1d & -0.000011 & -0.000014 & -0.000020 & -0.000054 & -0.000086 \\
2d & -0.000003 & -0.000006 & -0.000009 & -0.000019 & -0.000034 \\
3d & 0.000000  & -0.000001 & -0.000002 & -0.000004 & -0.000008 \\
4d & 0.000000  & -0.000001 & -0.000002 & -0.000003 & -0.000009 \\
5d & -0.000001 & -0.000003 & -0.000007 & -0.000019 & -0.000029 \\ 
6d & 0.000000  & 0.000000  & 0.000000  & -0.000001 & -0.000003 \\
7d & 0.000000  & -0.000001 & -0.000002 & -0.000004 & -0.000007 \\
\hline
1f & -0.000005 & -0.000008 & -0.000013 & -0.000051 & -0.000077 \\
2f & -0.000002 & -0.000004 & -0.000007 & -0.000018 & -0.000029 \\
3f & 0.000000  & -0.000001 & -0.000002 & -0.000003 & -0.000007 \\
4f & 0.000000  & 0.000000  & 0.000000  & 0.000002  & 0.000002  \\
5f & 0.000000  & 0.000000  & -0.000001 & -0.000001 & -0.000003 \\
\hline
1g & -0.000003 & -0.000006 & -0.000012 & -0.000030 & -0.000045 \\
2g & -0.000001 & -0.000002 & -0.000005 & -0.000011 & -0.000018 \\
3g & 0.000000  & -0.000001 & -0.000002 & -0.000004 & -0.000008 \\
4g & 0.000000  & 0.000000  & 0.000000  & 0.000000  & 0.000001  \\
\hline
1h & -0.000002 & -0.000004 & -0.000008 & -0.000018 & -0.000030 \\
2h & -0.000001 & -0.000002 & -0.000003 & -0.000008 & -0.000014 \\
3h & 0.000000  & -0.000001 & -0.000001 & -0.000003 & -0.000006 \\
\hline
1i & -0.000001 & -0.000003 & -0.000005 & -0.000013 & -0.000023 \\
2i & 0.000000  & -0.000001 & -0.000002 & -0.000006 & -0.000010 \\
\hline\hline
\end{tabular}
\end{table*}

We have thus established that for orbital basis sets --- where we \emph{can} make the comparison ---  pANO and ANO contractions of the same size are functionally largely equivalent in quality. This gives us the foundation we need to carry out pANO basis set optimizations in an F12 context, where true ANOs are not feasible.

\subsection{pANO-F12 contraction scheme}

We now proceed to explicitly correlated basis sets. In the original cc-pV$n$Z-F12 paper,\cite{peterson2008systematically} open-shell atomic MP2-F12 energy optimizations were found to lead to erratic behavior, and hence the optimizations for each element were carried out on an average of molecular closed-shell MP2-F12 energies for a small set of molecules for each chemical element. We employed the same diatomics and small polyatomics as Ref. \citenum{peterson2008systematically} in the present work.

As the primitive set, we started with a decontracted cc-pV5Z-F12 basis set\cite{jmlm261,jmlm269}.
 Again, in order not to `clutter' comparisons, for the relevant elements we at first retained the $sp$ contractions from cc-pV5Z-F12 verbatim, then optimized successive $1d$, $2d$, \ldots, $1f$, $2f$, \ldots, $1g$, $2g$, \ldots, and $1h$, $2h$ pANOs.
 For the `objective function' molecules of each element, we applied the full cc-pV5Z-F12 basis set to each non-target atom.
The auxiliary basis sets used were the MOLPRO 2024.1 defaults for the cc-pV5Z-F12 basis set, while the geminal exponent $\gamma$ was kept fixed at 1.4 following Ref.\cite{peterson2008systematically}. (A very recent paper\cite{Valeev2025slimmer} argues for even tighter $\gamma$ values in CCSD-F12.)

Next, to obtain a contraction scheme for the $p$ function, we copied $s$ and $d, f, g, h$ parts from cc-pV5Z-F12  as well as the first p contraction (that is, the SCF orbitals) and then optimized successive $2p$, $3p$, \ldots contractions. In order to elucidate the preferred contraction schemes,  we considered first-row elements, second-row elements, and hydrogen separately, evaluated the MP2-F12 total atomization energies (TAE$_e$ values, i.e. the sum of all bond energies) for all 200 entries in the W4-17 thermochemistry benchmark\cite{jmlm273}  and evaluated the RMSD with respect to cc-pV5Z-F12 results obtained with $\gamma$=1.2 as recommended by Peterson et al.\cite{jmlm261}. For comparison, the corresponding RMSDs were evaluated for standard cc-pV$n$Z-F12 basis sets, where $n=\{\mathrm{D,T,Q}\}$ with $\gamma$=\{0.9, 1.0, 1.0\}, respectively. 

At the final stage, we optimized pANO contractions of the $s$ functions from cc-pV5Z-F12 in the presence of the $pdfgh$ cc-pV5Z-F12 contraction and the SCF $s$ orbitals. These were then substituted into the already obtained pANO-$n$Z-F12 basis sets.

In order to avoid `dangling' diffuse $s$ and $p$ primitives, we re-contracted the $[2s1p]$ for first-row, and $[3s2p]$ for second row, SCF occupied orbitals using all aug-cc-pV6Z primitives\cite{dunning1996gaussian6, dunning2000gaussian7} of $s,p$ symmetries, including the most diffuse ones. (The said $sp$ primitives are, incidentally, the same as cc-pV5Z-F12\cite{peterson2008systematically,jmlm261}.) That is, they were contracted from $(17s11p)$ for the first-row elements and from $(22s15p)$ for the second-row elements.

\subsubsection{First-row elements: preferred contraction scheme}
To elucidate potential basis set structures for the first-row elements, species containing second-row elements were excluded from the W4-17 benchmark. For hydrogen-containing molecules, the standard cc-pV5Z-F12 basis set was used on hydrogen at this stage. The resulting Table \ref{tab:RMSD-V5Z-row1} presents the RMSD values obtained for the different contraction schemes.

{\scriptsize
\begin{table*}[!h]
\caption{RMSD (\kcalmol) from df-MP2-F12/cc-pV5Z-F12 for the first-row subset (128$^*$ species) of the W4-17 thermochemical benchmark\cite{jmlm273} as a function of pANO-F12 ($\gamma$=1.2) contraction scheme.}
\label{tab:RMSD-V5Z-row1}

\begin{tabular}{|l|l|l|l|l|l|l|l|}
\hline\hline
Contractions & 2p & 3p & 4p & 5p & 6p & 7p & 8p \\
\hline
s(V5Z-F12)                     & 24.289 & 22.286 & 21.864                       & 21.735                       & 21.917                       & 22.008                       & 22.022 \\
s(V5Z-F12) + 1d(pANO)          & 7.454  & 2.931  & {\color[HTML]{00B0F0} 1.992} & 1.708                        & 1.752                        & 1.788                        & 1.792  \\
s(V5Z-F12) + 2d(pANO)          &        & 2.672  & 1.494                        & {\color[HTML]{FF0000} 1.069} & 1.072                        & 1.096                        & 1.097  \\
s(V5Z-F12) + 2d+1f(pANO)       &        & 2.060  & 0.810                        & {\color[HTML]{00B0F0} 0.338} & 0.309                        & 0.311                        & 0.310  \\
s(V5Z-F12) + 3d(pANO)          &        &        & 1.506                        & 1.030                        & 1.002                        & 1.023                        & 1.023  \\
s(V5Z-F12) + 3d+1f(pANO)       &        &        & 0.802                        & 0.268                        & 0.230                        & 0.224                        & 0.221  \\
s(V5Z-F12) + 3d+2f(pANO)       &        &        & 0.915                        & 0.225                        & {\color[HTML]{FF0000} 0.155} & 0.154                        & 0.151  \\
s(V5Z-F12) + 3d+2f+1g(pANO)    &        &        & 0.995                        & 0.208                        & {\color[HTML]{00B0F0} 0.091} & 0.093                        & 0.089  \\
s(V5Z-F12) + 4d(pANO)          &        &        &                              & 1.030                        & 0.987                        & 0.999                        & 0.999  \\
s(V5Z-F12) + 4d+1f(pANO)       &        &        &                              & 0.236                        & 0.202                        & 0.193                        & 0.191  \\
s(V5Z-F12) + 4d+2f(pANO)       &        &        &                              & 0.192                        & 0.140                        & 0.135                        & 0.132  \\
s(V5Z-F12) + 4d+3f(pANO)       &        &        &                              & 0.202                        & 0.121                        & 0.120                        & 0.117  \\
s(V5Z-F12) + 4d+3f+1g(pANO)    &        &        &                              & 0.191                        & 0.061                        & 0.062                        & 0.059  \\
s(V5Z-F12) + 4d+3f+2g(pANO)    &        &        &                              & 0.222                        & 0.038                        & {\color[HTML]{FF0000} 0.037} & 0.031  \\
s(V5Z-F12) + 4d+3f+2g+1h(pANO) &        &        &                              & 0.238                        & 0.031                        & {\color[HTML]{00B0F0} 0.028} & 0.018 \\ 
\hline\hline
cc-pVDZ-F12; 5Z-F12 on H & \multicolumn{7}{c|}{1.251} \\
cc-pVTZ-F12; 5Z-F12 on H & \multicolumn{7}{c|}{0.220} \\
cc-pVQZ-F12; 5Z-F12 on H & \multicolumn{7}{c|}{0.037} \\
cc-pV5Z-F12 & \multicolumn{7}{c|}{REF} \\
\hline
\end{tabular}

$*$ Benzene was omitted because of basis set linear dependency issues.

\end{table*}
}

As expected, if we do not add \emph{any} angular correlation ($d$, $f$, $g$, $h$) functions, staggeringly high RMSD values are obtained. Obviously, at least a [1d] contraction is necessary, which in turn implies at least [4p]. Hence, the absolute minimum size is [4p1d] (which we shall denote pANO-DZ-F12-econ, the suffix being short for `economy'). Since adding the second $d$ contraction reduces RMSD further than adding the fifth $p$ contraction (0.5 kcal/mol vs 0.28 kcal/mol), there is little reason to use the [5p1d] contraction scheme. If we do start out with [2d] contractions, adding the fifth $p$ contraction decreases the RMSD value enough (0.43 kcal/mol) that it is starting to approach the [1d] $\rightarrow$ [2d] lowering with \textbf{four} $p$ contractions. We can thus rationalize [5p2d] as a pANO-DZ-F12-premium option.

Since the [2d] $\rightarrow$ [2d1f] improvement (about 0.7 kcal/mol) is comparable to the above [4p] $\rightarrow$ [5p] (about 0.5 kcal/mol), [5p2d1f] can be proposed as pANO-TZ-F12-econ. The further improvements achieved by adding the third $d$ and adding the second $f$ are comparable in magnitude; since they are also comparable to those from adding the sixth $p$, we propose [6p3d2f] as pANO-TZ-F12-premium.

Logically, the most appropriate next step beyond pANO-TZ-F12 would be to add the first $g$ contraction. Accordingly, we propose [6p3d2f1g] as the pANO-QZ-F12-econ basis set.  Addition of an extra $p$ contraction does not materially affect RMSD. In contrast, adding one more $d$ contraction does slightly reduce RMSD, while simultaneously adding one $d$ and one $f$ contractions does so even more. Simply adding $[1d1f1g]$ contractions all at once significantly reduces the RMSD value (by almost 60\%) and thus we can propose [6p4d3f2g] as pANO-QZ-F12-premium. Alternatively, one might propose [7p4d3f2g] as pANO-QZ-F12-premium, which is only negligibly different (ca. 0.001 kcal/mol) from  [6p4d3f2g], but it would establish a pattern that pANO-$n$Z-F12-premium is pANO-$n$Z-F12-econ plus one of each angular momentum.
Laying down the complementary pattern that pANO-(n+1)Z-F12-econ is pANO-$n$Z-F12-premium plus one extra angular momentum, we can thus propose [7p4d3f2g1h] as pANO-5Z-F12. 

\subsubsection{Hydrogen contraction scheme}

Both the large cc-pV5Z-F12 basis set\cite{jmlm261} and its variant for hydrogen cc-pV5Z-F12rev2 with an additional layer of $dfg$ functions\cite{jmlm261} were considered for optimization of hydrogen's contraction scheme, but in the end only the latter was retained. Once again, in order not to `muddle' comparisons, we at first copied the $s$ part from the corresponding basis set and then optimized successive $1p$, $2p$, \ldots, $1d$, $2d$, \ldots, $1f$, $2f$, \ldots, and $1g$, $2g$ pANOs. To obtain $s$ contractions, we retained the whole 
$pdfgh$ part as well as the contracted SCF $s$ orbital for hydrogen atom, and then optimized $1s$, $2s$, \ldots on top. Obviously, for this purpose only the hydrogen-containing subset of W4-17 was considered; in this set of runs, the full cc-pV5Z-F12 basis set was used on all nonhydrogen atoms. Table \ref{tab:RMSD-Hygrogen-V5Z-F12rev2} presents relevant RMSD values.

{\scriptsize
\begin{table*}[!h]
\caption{RMSD (\kcalmol) from df-MP2-F12/cc-pV5Z-F12rev2\cite{jmlm261} for the 126 hydrogen-containing species of the W4-17 thermochemical benchmark\cite{jmlm273} as a function of pANO-F12 ($\gamma$=1.2) contraction scheme for hydrogen. The primitives from cc-pV5Z-F12rev2\cite{jmlm261} were used.}
\label{tab:RMSD-Hygrogen-V5Z-F12rev2}
\begin{tabular}{|l|l|l|l|l|l|l|l|}
\hline\hline
Contractions & 2s & 3s & 4s & 5s & 6s & 7s & 8s \\
\hline
$s_i$  & 0.406 & 0.373 & 0.308 & 0.299 & 0.297 & 0.295 & 0.287 \\

$s_i$ + 1p(pANO) & 0.140 & 0.122 & {\color[HTML]{00B0F0} 0.111} & 0.107 & 0.106 & 0.105 & 0.101 \\

$s_i$ + 2p(pANO) &  & 0.100 & 0.094 & {\color[HTML]{FF0000} 0.092} & 0.091 & 0.090 & 0.083 \\
$s_i$ + 2p+1d(pANO) &  & 0.071 & 0.068 & {\color[HTML]{00B0F0} 0.067} & 0.067 & 0.066 & 0.064 \\

$s_i$ + 3p(pANO) &  &  & 0.086 & 0.084 & 0.083 & 0.082 & 0.075 \\
$s_i$ + 3p+1d(pANO) &  &  & 0.060 & 0.059 & 0.059 & 0.057 & 0.055 \\
$s_i$ + 3p+2d(pANO) &  &  & 0.046 & 0.046 & {\color[HTML]{FF0000} 0.045} & 0.044 & 0.042 \\
$s_i$ + 3p+2d+1f(pANO) &  &  & 0.039 & 0.038 & {\color[HTML]{00B0F0} 0.038} & 0.037 & 0.035 \\

$s_i$ + 4p(pANO) &  &  &  & 0.078 & 0.077 & 0.076 & 0.069 \\
$s_i$ + 4p+1d(pANO) &  &  &  & 0.054 & 0.054 & 0.052 & 0.050 \\
$s_i$ + 4p+2d(pANO) &  &  &  & 0.039 & 0.039 & 0.038 & 0.035 \\
$s_i$ + 4p+2d+1f(pANO) &  &  &  & 0.031 & 0.031 & 0.030 & 0.028 \\

$s_i$ + 4p+3d(pANO) &  &  &  & 0.033 & 0.033 & 0.032 & 0.030 \\
$s_i$ + 4p+3d+1f(pANO) &  &  &  & 0.026 & 0.025 & 0.024 & 0.023 \\
$s_i$ + 4p+3d+2f(pANO) &  &  &  & 0.022 & 0.021 & {\color[HTML]{FF0000} 0.020} & 0.019 \\
$s_i$ + 4p+3d+2f+1g(pANO) &  &  &  & 0.017 & 0.017 & {\color[HTML]{00B0F0} 0.016} & 0.015 \\
\hline\hline
5Z-F12 on non H; cc-pVDZ-F12rev2 & \multicolumn{7}{c|}{0.194} \\
5Z-F12 on non H; cc-pVTZ-F12rev2 & \multicolumn{7}{c|}{0.068} \\
5Z-F12 on non H; cc-pVQZ-F12rev2 & \multicolumn{7}{c|}{0.017} \\
5Z-F12 on non H; cc-pV5Z-F12rev2 & \multicolumn{7}{c|}{REF} \\
\hline

\end{tabular}

\end{table*}
}

We thus finally obtain the following contraction schemes for hydrogen: [4s1p] for pANO-DZ-F12-econ, [5s2p] for pANO-DZ-F12-premium, [5s2p1d] for pANO-TZ-F12-econ, [6s3p2d] for pANO-TZ-F12-premium, [6s3p2d1f] for pANO-QZ-F12-econ, [7s4p3d2f] for pANO-QZ-F12-premium, and [7s4p3d2f1g] for pANO-5Z-F12.

\subsubsection{Second-row elements contraction scheme}  
To survey potential contraction schemes for the second-row elements, only the subset of W4-17 that contains second-row elements was considered. Several species were excluded as outliers that drive up RMSD to the point of obscuring `signal' from the changes in the other species: those include \ce{S3} and \ce{S4} which are beset by strong static correlation and open-shell \ce{S2} which has moderate static correlation, as well as \ce{P4} and \ce{P2} which have notoriously slow basis set convergence\cite{Persson1997P4}. To the first-row elements and hydrogen, the full cc-pV5Z-F12 basis set was assigned during the optimization. The resulting Table \ref{tab:RMSD-V5Z-row2} presents the RMSD values obtained for the different contraction schemes.
 
{\scriptsize
\begin{table*}[!h]
\caption{RMSD (\kcalmol) from MP2-F12/cc-pV5Z-F12 for the second-row subset (71 species) of the W4-17 thermochemical benchmark\cite{jmlm273} as a function of pANO-F12 ($\gamma$=1.2) contraction scheme.}
\label{tab:RMSD-V5Z-row2}

\begin{tabular}{|l|l|l|l|l|l|l|l|}
\hline\hline
Contractions & 3p & 4p & 5p & 6p & 7p & 8p & 9p \\
\hline

s(V5Z-F12)                     & 27.813 & 29.501 & 29.785                       & 29.947                       & 30.048                       & 30.098                       & 30.130 \\
s(V5Z-F12) + 1d(pANO)          & 4.870  & 4.707  & 4.694                        & 4.692                        & 4.734                        & 4.744                        & 4.758  \\
s(V5Z-F12) + 2d(pANO)          &        & 4.166  & {\color[HTML]{00B0F0} 4.079} & 4.071                        & 4.112                        & 4.122                        & 4.128  \\
s(V5Z-F12) + 2d+1f(pANO)       &        & 0.919  & 0.797                        & 0.755                        & 0.760                        & 0.765                        & 0.766  \\
s(V5Z-F12) + 3d(pANO)          &        &        & 4.070                        & {\color[HTML]{FF0000} 4.059} & 4.094                        & 4.103                        & 4.105  \\
s(V5Z-F12) + 3d+1f(pANO)       &        &        & 0.733                        & {\color[HTML]{00B0F0} 0.686} & 0.689                        & 0.693                        & 0.693  \\
s(V5Z-F12) + 4d(pANO)          &        &        &                              & 4.085                        & 4.119                        & 4.129                        & 4.131  \\
s(V5Z-F12) + 4d+1f(pANO)       &        &        &                              & 0.691                        & 0.692                        & 0.696                        & 0.696  \\
s(V5Z-F12) + 4d+2f(pANO)       &        &        &                              & 0.524                        & {\color[HTML]{FF0000} 0.508} & 0.511                        & 0.510  \\
s(V5Z-F12) + 4d+2f+1g(pANO)    &        &        &                              & 0.148                        & {\color[HTML]{00B0F0} 0.128} & 0.129                        & 0.129  \\
s(V5Z-F12) + 5d(pANO)          &        &        &                              &                              & 4.147                        & 4.157                        & 4.159  \\
s(V5Z-F12) + 5d+1f(pANO)       &        &        &                              &                              & 0.698                        & 0.702                        & 0.702  \\
s(V5Z-F12) + 5d+2f(pANO)       &        &        &                              &                              & 0.511                        & 0.514                        & 0.513  \\
s(V5Z-F12) + 5d+2f+1g(pANO)    &        &        &                              &                              & 0.125                        & 0.126                        & 0.126  \\
s(V5Z-F12) + 5d+3f(pANO)       &        &        &                              &                              & 0.497                        & 0.498                        & 0.498  \\
s(V5Z-F12) + 5d+3f+1g(pANO)    &        &        &                              &                              & 0.106                        & 0.106                        & 0.105  \\
s(V5Z-F12) + 5d+3f+2g(pANO)    &        &        &                              &                              & 0.081                        & {\color[HTML]{FF0000} 0.080} & 0.079  \\
s(V5Z-F12) + 5d+3f+2g+1h(pANO) &        &        &                              &                              & 0.051                        & {\color[HTML]{00B0F0} 0.050} & 0.049  \\

\hline\hline
cc-pVDZ-F12; 5Z-F12 on H, B--F & \multicolumn{7}{c|}{4.297} \\
cc-pVTZ-F12; 5Z-F12 on H, B--F & \multicolumn{7}{c|}{0.604} \\
cc-pVQZ-F12; 5Z-F12 on H, B--F & \multicolumn{7}{c|}{0.124} \\
cc-pV5Z-F12 on H, B--F, Al--Cl & \multicolumn{7}{c|}{REF} \\
\hline
cc-pVTZ-F12(no F); 5Z-F12 on H, B--F & \multicolumn{7}{c|}{4.097} \\
5Z-F12, aug-cc-pV(5+d)Z-JK on Al--Cl & \multicolumn{7}{c|}{0.120} \\
\hline

\end{tabular}

\end{table*}
}

All results are summarized in Table \ref{tab:RMSD-V5Z-row2}. 
(The first two $p$ contractions are obviously the SCF atomic orbitals.)
Here it is better to use [2d] contractions instead of only [1d] as we did for first-row elements, because [1d] $\rightarrow$ [2d] makes for a significant reduction of RMSD. For [2d] we need at least [5p] contractions, which turns us to a minimum pANO-DZ-F12-econ size for second-row elements of [5p2d]. Again similarly to the first-row case, adding the sixth $p$ causes a similar reduction of RMSD as adding the third $d$ does. Thus, [6p3d] will be pANO-DZ-F12-premium.

Adding the first $f$, we proceed to the TZ contraction scheme. [6p3d1f] has the best RMSD reduction among all [3d1f], thus establishing [6p3d1f] is pANO-TZ-F12-econ. Simultaneous addition of $d$ and $f$ functions is reasonable, leading to [4d2f] for pANO-TZ-premium. [7p4d2f] has the minimum RMSD among all [4d2f], finally begetting [7p4d2f] as pANO-TZ-F12-premium.

Next, the next angular momentum $g$ to our pANO-TZ-F12-premium should be added to establish pANO-QZ-F12-econ as [7p4d2f1g]. Again, in order to obtain pANO-QZ-F12-premium, we need to add one contraction of each angular momentum, obtaining [8p5d3f2g]. Following the pattern established above, that pANO-(n+1)Z-F12-econ amounts to pANO-$n$Z-F12-premium plus one the next angular momentum, we can propose [8p5d3f2g1h] as pANO-5Z-F12. 

Comparison of the convergence patterns of first- and second-row elements (Tables \ref{tab:RMSD-V5Z-row1} and \ref{tab:RMSD-V5Z-row2}) reveals something intriguing that transcends the present study: apparently, F12 basis set convergence is so much slower for the second row than for the first row, that an $spdf$ basis set is needed for the 2nd row to meet the performance level that can be achieved with just $spd$, and likewise for second-row $spdfg$ vs. first-row $spdf$. This finding is not specific to pANO-F12 but applies also to cc-pV$n$Z-F12; see, e.g., Barman et al.\cite{jmlm340}

\subsection{Contraction scheme for the \textit{s} functions}
Having established [p,d,f,g,h] contraction schemes for pANO-$n$Z-F12 basis set family on top of the full complement of $s$ functions from cc-pV5Z-F12, it would appear to make sense to contract down the $s$ functions for nonhydrogen elements too, even if the gain in computational efficiency thus achieved is much smaller than from contracting away $p$ and especially $d$, $f$, \ldots functions. As already stated above, these contractions were obtained from the full cc-pV5Z-F12 $pdfgh$ set and the relevant SCF atomic orbitals of $s$ symmetry.
Table \ref{tab:RMSD-V5Z-row1-Scontractions} for the first-row elements, and Table \ref{tab:RMSD-V5Z-row2-Scontractions} for the second row elements, display the dependence on the number of $s$ contractions of the RMSD for the W4-17
thermochemical benchmark. 

{\scriptsize
\begin{table*}[!h]
\caption{RMSD (\kcalmol) from df-MP2-F12/cc-pV5Z-F12rev2\cite{jmlm261} for the first-row subset (128$^*$ species) of the W4-17 thermochemical benchmark\cite{jmlm273}.}
\label{tab:RMSD-V5Z-row1-Scontractions}

\begin{tabular}{|l|l|l|l|l|l|l|}
\hline\hline
Contractions & 3s & 4s & 5s & 6s & 7s & 8s  \\
\hline



pANO-DZ-econ [4p1d]$^a$     & 1.261 & {\color[HTML]{00B0F0} 1.315} & 1.510                        & 1.386                        & 1.341 & 1.321 \\
pANO-DZ-prem [5p2d]$^a$      & 0.936 & 0.772                        & {\color[HTML]{FF0000} 1.115} & 1.061                        & 1.082 & 1.109 \\ 

pANO-TZ-econ [5p2d1f]$^a$    & 1.693 & 0.445                        & {\color[HTML]{00B0F0} 0.448} & 0.420                        & 0.428 & 0.458 \\
pANO-TZ-prem [6p3d2f]$^a$    & 2.084 & 0.541                        & 0.172                        & {\color[HTML]{FF0000} 0.162} & 0.181 & 0.230 \\

pANO-QZ-econ [6p3d2f1g] & 1.740 & 0.585                        & 0.094                        & {\color[HTML]{00B0F0} 0.073} & 0.086                        & 0.123                        \\
pANO-QZ-prem [7p4d3f2g] & 1.910 & 0.660                        & 0.109                        & 0.084                        & {\color[HTML]{FF0000} 0.078} & 0.080                        \\

pANO-5Z [7p4d3f2g1h]    & 3.020 & 0.757                        & 0.143                        & 0.102                        & 0.066                        & {\color[HTML]{00B0F0} 0.028} \\
\hline
&\multicolumn{6}{c}{Rel. to 8s same row}\\					
DZ-econ [4p1d]       & 1.129 & {\color[HTML]{00B0F0}0.491} & 0.320 & 0.160 & 0.119 & REF \\
DZ-prem [5p2d]       & 1.631 & 0.595 & {\color[HTML]{FF0000}0.103} & 0.082 & 0.078 & REF \\
TZ-econ [5p2d1f]     & 2.023 & 0.590 & {\color[HTML]{00B0F0}0.108} & 0.086 & 0.075 & REF \\
TZ-prem [6p3d2f]     & 2.326 & 0.712 & 0.126 & {\color[HTML]{FF0000}0.109} & 0.090 & REF \\
QZ-econ [6p3d2f1g]   & 1.876 & 0.688 & 0.132 & {\color[HTML]{00B0F0}0.111} & 0.092 & REF \\
QZ-prem [7p4d3f2g]   & 2.011 & 0.737 & 0.151 & 0.122 & {\color[HTML]{FF0000}0.097} & REF \\
5Z [7p4d3f2g1h]      & 3.106 & 0.778 & 0.148 & 0.107 & 0.074 & {\color[HTML]{00B0F0}REF} \\
\hline\hline
\end{tabular}

$*$ Benzene was omitted because of basis set linear dependency issues. \\
$^a$ aug-cc-pVQZ-F12/MP2FIT (DF), aug-cc-pVQZ-F12/JKFIT (JK), and aug-cc-pwCV5Z/MP2FIT (CABS) were used as auxiliary basis sets.

\end{table*}
}

{\scriptsize
\begin{table*}[!h]
\caption{RMSD (\kcalmol) from df-MP2-F12/cc-pV5Z-F12rev2 for the second-row subset (71 species) of the W4-17 thermochemical benchmark\cite{jmlm273}.}
\label{tab:RMSD-V5Z-row2-Scontractions}

\begin{tabular}{|l|l|l|l|l|l|l|}
\hline\hline
Contractions & 4s & 5s & 6s & 7s & 8s & 9s \\
\hline



pANO-DZ-econ [5p2d]$^a$     & 0.798 & {\color[HTML]{00B0F0} 0.769} & 0.851                        & 0.829                        & 0.827 & 0.802 \\
pANO-DZ-prem [6p3d]$^a$     & 0.784 & 0.783                        & {\color[HTML]{FF0000} 0.857} & 0.874                        & 0.874 & 0.866 \\

pANO-TZ-econ [6p3d1f]$^a$   & 0.368 & 0.326                        & {\color[HTML]{00B0F0} 0.485} & 0.514                        & 0.508 & 0.490 \\
pANO-TZ-prem [7p4d2f]$^a$   & 0.199 & 0.225                        & 0.213                        & {\color[HTML]{FF0000} 0.241} & 0.243 & 0.240 \\

pANO-QZ-econ [7p4d2f1g] & 0.326 & 0.277                        & 0.134                        & {\color[HTML]{00B0F0} 0.148} & 0.157                        & 0.159 \\
pANO-QZ-prem [8p5d3f2g] & 0.461 & 0.315                        & 0.105                        & 0.093                        & {\color[HTML]{FF0000} 0.106} & 0.113 \\

pANO-5Z [8p5d3f2g1h]    & 0.261 & 0.230                        & 0.085                        & 0.064                        & {\color[HTML]{00B0F0} 0.064} & 0.066 \\

\hline
&\multicolumn{6}{c}{Rel. to 9s same row}\\
DZ-econ [5p2d]       & 0.153 & {\color[HTML]{00B0F0}0.144} & 0.170 & 0.083 & 0.074 & REF \\
DZ-prem [6p3d]       & 0.185 & 0.285 & {\color[HTML]{FF0000}0.096} & 0.061 & 0.050 & REF \\
TZ-econ [6p3d1f]     & 0.168 & 0.212 & {\color[HTML]{00B0F0}0.071} & 0.039 & 0.027 & REF \\
TZ-prem [7p4d2f]     & 0.270 & 0.301 & 0.087 & {\color[HTML]{FF0000}0.019} & 0.013 & REF \\
QZ-econ [7p4d2f1g]   & 0.378 & 0.317 & 0.098 & {\color[HTML]{00B0F0}0.037} & 0.007 & REF \\
QZ-prem [8p5d3f2g]   & 0.520 & 0.370 & 0.118 & 0.049 & {\color[HTML]{FF0000}0.012} & REF \\
5Z [8p5d3f2g1h]      & 0.276 & 0.246 & 0.078 & 0.012 & {\color[HTML]{00B0F0}0.006} & REF \\
\hline\hline
\end{tabular}

$^a$ aug-cc-pV5Z-F12/MP2FIT (DF), aug-cc-pV(5+d)Z-F12/JKFIT (JK), and aug-cc-pwCV5Z/MP2FIT (CABS) were used as auxiliary basis sets.

\end{table*}
}

  Because the absolute RMSDs in Tables~\ref{tab:RMSD-V5Z-row1-Scontractions} and \ref{tab:RMSD-V5Z-row2-Scontractions} are affected by error compensation with the residual incompleteness in the higher angular momenta (which is why they are not monotonic in the number of $s$ contractions), the $s$ contraction was instead selected on the basis of the RMS difference from the maximally $s$-decontracted basis with the same higher-$l$ pattern (lower blocks of the tables), subject to the constraint $n_s \geq n_p$. Otherwise, we proceeded by analogy with a scree plot\cite{Cattell1966scree} (as in principal component analysis): we sought the `elbow'/`knee' beyond which further $s$ decontraction yields diminishing returns. In practice, the constraint is always at least as restrictive as this criterion, so that $n_s = n_p$ for all econ and prem sets; the only exception is the first-row pANO-5Z, for which 8$s$ was retained given its role as a reference.

The resulting contraction patterns for the complete pANO-F12 family are summarized in Table~\ref{tab:ContractionSizes}, together with those of the corresponding cc-pV$n$Z-F12\cite{peterson2008systematically,jmlm261} and, for hydrogen, cc-pV$n$Z-F12rev2\cite{jmlm261} basis sets. For B--F and Al--Cl, the pANO-$n$Z-F12-prem sets ($n$=D,T,Q) have exactly the same contracted shell structure as cc-pV$n$Z-F12, differing only in the contraction coefficients; for hydrogen, they correspond to cc-pV$n$Z-F12rev2\cite{jmlm261} plus two additional $s$ contractions. The pANO-$n$Z-F12-econ sets are about 25--30\% smaller than cc-pV$n$Z-F12 for the heavy atoms. pANO-5Z-F12 follows the economy construction and is about 25\% smaller than cc-pV5Z-F12. We note that all pANO-F12 sets are contracted from the cc-pV5Z-F12 (for H, cc-pV5Z-F12rev2) primitive set, so that particularly at the DZ and TZ levels, their primitive counts are considerably larger than those of cc-pV$n$Z-F12. In canonical CCSD-F12 calculations, this is not a concern, as the integral evaluation time will typically be a small fraction of the wall clock execution time.

\begin{table*}
\caption{Contraction patterns and numbers of contracted basis functions $N_{\rm cont.}$ (spherical harmonics) per atom for the cc-pV$n$Z-F12 and pANO-$n$Z-F12 basis set families.}
\label{tab:ContractionSizes}
\small
\begin{tabular}{l|l@{\hspace{2pt}}r|l@{\hspace{2pt}}r|l@{\hspace{2pt}}r|l@{\hspace{2pt}}r}
\hline
 & \multicolumn{2}{c|}{DZ} & \multicolumn{2}{c|}{TZ} & \multicolumn{2}{c|}{QZ} & \multicolumn{2}{c}{5Z} \\
 & contr. & $N_{\rm cont.}$ & contr. & $N_{\rm cont.}$ & contr. & $N_{\rm cont.}$ & contr. & $N_{\rm cont.}$ \\
\hline
\multicolumn{9}{l}{H} \\
cc-pV$n$Z-F12     & [3s2p] & 9  & [4s3p1d] & 18 & [5s4p2d1f] & 34 & [6s5p3d2f1g] & 59 \\
cc-pV$n$Z-F12rev2 & [3s2p] & 9  & [4s3p2d] & 23 & [5s4p3d2f] & 46 & [6s5p4d3f2g] & 80 \\
pANO-econ         & [4s1p] & 7  & [5s2p1d] & 16 & [6s3p2d1f] & 32 & [7s4p3d2f1g]$^a$ & 57 \\
pANO-prem         & [5s2p] & 11 & [6s3p2d] & 25 & [7s4p3d2f] & 48 & \multicolumn{2}{c}{---} \\
\hline
\multicolumn{9}{l}{B--F} \\
cc-pV$n$Z-F12     & [5s5p2d] & 30 & [6s6p3d2f] & 53 & [7s7p4d3f2g] & 87 & [8s8p5d4f3g2h] & 134 \\
pANO-econ         & [4s4p1d] & 21 & [5s5p2d1f] & 37 & [6s6p3d2f1g] & 62 & [8s7p4d3f2g1h]$^a$ & 99 \\
pANO-prem         & [5s5p2d] & 30 & [6s6p3d2f] & 53 & [7s7p4d3f2g] & 87 & \multicolumn{2}{c}{---} \\
\hline
\multicolumn{9}{l}{Al--Cl} \\
cc-pV$n$Z-F12     & [6s6p3d] & 39 & [7s7p4d2f] & 62 & [8s8p5d3f2g] & 96 & [9s8p6d4f3g2h] & 140 \\
pANO-econ         & [5s5p2d] & 30 & [6s6p3d1f] & 46 & [7s7p4d2f1g] & 71 & [8s8p5d3f2g1h]$^a$ & 107 \\
pANO-prem         & [6s6p3d] & 39 & [7s7p4d2f] & 62 & [8s8p5d3f2g] & 96 & \multicolumn{2}{c}{---} \\
\hline
\end{tabular}

\smallskip
\footnotesize
$^a$ pANO-5Z-F12, constructed as pANO-QZ-F12-prem plus one contraction of the next angular momentum; there is no separate economy/premium distinction at the 5Z level.
\end{table*}

\subsection{Diffuse functions and revisiting S66 and W4-17}

It has been known since at least the mid-1980s\cite{Frisch1986HBond} that adding diffuse functions (a.k.a., `anion functions') to the basis set is highly beneficial for noncovalent interactions, both in orbital calculations and in F12 calculations. (They are obviously essential for evaluating electron affinities.) For that latter purpose, the aug-cc-pV$n$Z-F12 basis sets were developed in our group\cite{jmlm276}: since the underlying cc-pV$n$Z-F12 already contains diffuse functions for the valence angular momenta, single $df\ldots$ primitives were added and their exponents optimized for the total energies of atomic anions. (As nitrogen and neon do not have bound anions, their diffuse function exponents were obtained by cubic interpolation in the atomic number Z, paralleling what was done in the original aug-cc-pV$n$Z paper\cite{Kendall1992} in the orbital-basis-set context.) 

Adding diffuse functions to a contracted ANO basis set creates a conundrum. Three approaches are possible:
\begin{itemize}
    \item Simply adding the next natural orbital for each angular momentum, as was done by Neese and Valeev\cite{NeeseValeev2010} --- without any guarantee that these additional NOs in fact do fortify the diffuse range of the wave function. (In practice, they do not.)
    \item The Widmark-Malmqvist-Roos approach,\cite{Widmark1990} which consists of averaging density matrices of the neutral atom, cation, anion, and neutral atom in an electric field, then diagonalizing. This is quite awkward to generalize to the present pANO framework.
    \item Simply optimizing individual primitives on top of a fixed contracted pANO-F12 basis set, as already hinted at in the second Alml\"of-Taylor paper\cite{AlmlofTaylorPart2}.
\end{itemize}

We availed ourselves of the latter option, and optimized $d$, $f$, \ldots (for hydrogen, also $p$) functions along the lines of Ref. \citenum{jmlm276} (including the cubic interpolation for nitrogen).    For the $sp$ functions of the first- and second-row elements, we decontracted the most diffuse primitive of the underlying basis set, which itself was optimized for the anion Hartree-Fock energy (in a practice going back at least to Ref.~\citenum{Kendall1992}); for hydrogen, the most diffuse $s$ exponent (0.0189) of aug-cc-pV6Z\cite{dunning1996gaussian6} was added.
 The resulting diffuse exponents can be found in the SI (on pages 2--3), and permit creation of aug-pANO-nZ-econ and aug-pANO-nZ-prem basis sets, respectively. 

The importance of diffuse functions for electron affinities can be seen in Table \ref{tab:EA-CCSD-F12b}, especially for hydrogen and boron. Of course, if the properties of interest were electron affinities, then diffuse functions would be essential in either basis set family.   

{\scriptsize
\begin{table*}[!h]
\caption{Electron affinities (eV) for separate species at the CCSD-F12b level of theory.}
\label{tab:EA-CCSD-F12b}

\begin{tabular}{|l|l|l|l|l|l|l|l|l|l|}
\hline\hline
Atom & DZ-econ & DZ-prem & TZ-econ & TZ-prem & QZ-econ & QZ-prem & 5Z & Ref$^a$ \\
\hline

 & \multicolumn{7}{l|}{non-augmented basis set family} & \\
\hline

H  & 0.32752  & 0.40582  & 0.42617  & 0.44953 & 0.44527 & 0.47913 & 0.49536 &  0.75389 \\
B  & -0.12689 & -0.03075 & -0.03262 & 0.01507 & 0.01533 & 0.04606 & 0.04646 &  0.18746 \\
C  & 0.86274  & 0.99518  & 1.00656  & 1.05682 & 1.05784 & 1.13380 & 1.14318 &  1.17297 \\
O  & 0.94532  & 1.09393  & 1.12990  & 1.23158 & 1.22852 & 1.24667 & 1.25564 &  1.28471 \\
F  & 2.89102  & 3.01287  & 2.97672  & 3.20160 & 3.20544 & 3.21858 & 3.22307 &  3.24579 \\
Al & 0.27134  & 0.31804  & 0.32508  & 0.33955 & 0.32993 & 0.35122 & 0.35634 &  0.38645 \\
Si & 1.22726  & 1.27801  & 1.29916  & 1.32421 & 1.32297 & 1.32565 & 1.33426 &  1.35085 \\
P  & 0.32105  & 0.39020  & 0.48293  & 0.51653 & 0.52836 & 0.59349 & 0.60556 &  0.64860 \\
S  & 1.74527  & 1.80194  & 1.85578  & 1.94005 & 1.95495 & 1.96999 & 1.97070 &  1.99113 \\
Cl & 3.41503  & 3.45764  & 3.45200  & 3.50527 & 3.51204 & 3.52611 & 3.53616 &  3.56195 \\
\textbf{RMSD:} & \textbf{0.289} & \textbf{0.204} & \textbf{0.186} & \textbf{0.129} & \textbf{0.129} & \textbf{0.103} & \textbf{0.096} & \\

\hline

 & \multicolumn{7}{l|}{augmented basis set family} & \\

\hline

H  & 0.74793 & 0.75148 & 0.75067 & 0.75158 & 0.75047 & 0.75197 & 0.75232 & 0.75389 \\
B  & 0.16853 & 0.17535 & 0.17935 & 0.17942 & 0.17932 & 0.18613 & 0.18586 & 0.18746 \\
C  & 1.15117 & 1.15483 & 1.16405 & 1.16853 & 1.16889 & 1.17302 & 1.17293 & 1.17297 \\
O  & 1.20430 & 1.23115 & 1.26519 & 1.27966 & 1.28372 & 1.28596 & 1.28720 & 1.28471 \\
F  & 3.17166 & 3.19762 & 3.21181 & 3.23708 & 3.24454 & 3.24628 & 3.24751 & 3.24579 \\
Al & 0.37043 & 0.37150 & 0.38734 & 0.38680 & 0.38683 & 0.38655 & 0.38632 & 0.38645 \\
Si & 1.31775 & 1.31778 & 1.34808 & 1.35135 & 1.35233 & 1.35034 & 1.35093 & 1.35085 \\
P  & 0.51027 & 0.51997 & 0.61354 & 0.61907 & 0.63563 & 0.64371 & 0.64567 & 0.64860 \\
S  & 1.89470 & 1.90235 & 1.95356 & 1.96685 & 1.98979 & 1.99088 & 1.99168 & 1.99113 \\
Cl & 3.52250 & 3.52320 & 3.52418 & 3.53057 & 3.56312 & 3.56183 & 3.56353 & 3.56195 \\
\textbf{RMSD:} & \textbf{0.066} & \textbf{0.057} & \textbf{0.024} & \textbf{0.016} & \textbf{0.005} & \textbf{0.002} & \textbf{0.002} & \\

\hline

\end{tabular}

$^a$ CCSD(F12*)/d-aug-cc-pwCV5Z results from Table III of Ref.\cite{jmlm276}

\end{table*}
}

Table \ref{tab:NCI-RMSDs-results} summarizes NCI results for the S66 benchmark\cite{rezac2011s66} at the PNO-LCCSD-F12 level with Tight DOMOPT settings. As explained at length by K\"ohn and Tew\cite{KohnTew2010CuspI,jmlm285}, the F12a method is fatally flawed and should not be used for production calculations; for water clusters, we found\cite{jmlm272} significant F12b-F12a differences even for basis sets as large as quintuple zeta (5Z).

As can be seen in the said Table \ref{tab:NCI-RMSDs-results}, adding the diffuse functions markedly reduces the counterpoise-corrected RMSD from the complete basis set limit. If \emph{circa} 0.1 kcal/mol accuracy is enough, then even aug-pANO-DZ-econ will do the job in conjunction with a CP correction, as will aug-pANO-DZ-prem in the absence of CP correction. 

\begin{table}[!htbp]
\caption{RMSD (in \kcalmol) for S66 benchmark obtained by the new pANO-$n$Z-F12 basis set family at the CCSD-F12b level and at the PNO-LCCSD-F12b level with DOMOPT=Tight threshold setting.}
\label{tab:NCI-RMSDs-results}
\scriptsize
\setlength{\tabcolsep}{3pt}
\begin{tabular}{|r|l|l|}
\hline

Basis  & RMSD CP & RMSD raw \\
\hline

\hline \hline
PNO-LCCSD-F12b/pANO-DZ-econ & 0.398 & 0.644 \\
CCSD-F12b/pANO-DZ-econ & 0.410 & 0.310 \\

PNO-LCCSD-F12b/pANO-TZ-econ & 0.159 & 0.294 \\
PNO-LCCSD-F12b/pANO-QZ-econ & 0.051 & 0.126 \\
\hline

PNO-LCCSD-F12b/pANO-DZ-prem & 0.164 & 0.290 \\
CCSD-F12b/pANO-DZ-prem & 0.164 & 0.199 \\

PNO-LCCSD-F12b/pANO-TZ-prem & 0.059 & 0.152 \\
PNO-LCCSD-F12b/pANO-QZ-prem & 0.011 & 0.055 \\

\hline

PNO-LCCSD-F12b/aug-pANO-DZ-econ     & 0.096 & 0.461 \\
CCSD-F12b/aug-pANO-DZ-econ & 0.154 & 0.409 \\

PNO-LCCSD-F12b/aug-pANO-TZ-econ & 0.020 & 0.113 \\
PNO-LCCSD-F12b/aug-pANO-QZ-econ & 0.009 & 0.069 \\

\hline

PNO-LCCSD-F12b/aug-pANO-DZ-prem     & 0.050 & 0.122 \\
CCSD-F12b/aug-pANO-DZ-prem & 0.101 & 0.160 \\

PNO-LCCSD-F12b/aug-pANO-TZ-prem & 0.014 & 0.097 \\  
PNO-LCCSD-F12b/aug-pANO-QZ-prem & 0.010 & 0.021 \\ 
\hline

PNO-LCCSD-F12b/cc-pVDZ-F12 & 0.098 & 0.096 \\
CCSD-F12b/cc-pVDZ-F12 & 0.153 & 0.143 \\

PNO-LCCSD-F12b/cc-pVTZ-F12 & 0.042 & 0.031 \\
CCSD-F12b/cc-pVTZ-F12 & 0.044 & 0.074 \\

PNO-LCCSD-F12b/cc-pVQZ-F12 & 0.012 & 0.017 \\ 
\hline

PNO-LCCSD-F12b/aug-pVDZ-F12 & 0.034 & 0.071 \\ 
CCSD-F12b/aug-pVDZ-F12 & 0.105 & 0.197 \\ 

PNO-LCCSD-F12b/aug-pVTZ-F12$^a$ & 0.013 & 0.020 \\ 
PNO-LCCSD-F12b/aug-pVQZ-F12$^a$ & 0.010 & 0.014 \\ 
\hline
PNO-LCCSD-F12b/pANO-5Z-F12 & REF & 0.044 \\ 
\hline
\end{tabular}

Geminal exponents $\gamma$=0.9, 1.0, 1.0, and 1.2 were used for DZ, TZ, QZ, and 5Z basis sets, respectively, following Refs.~\citenum{Hill2009_extrap_MP2_CCSD_F12,jmlm261}.

$^a$ Systems containing benzene are excluded.

\end{table}

At the final stage, an augmented pANO-$n$Z-F12 basis set has been assessed on W4-17 thermochemistry benchmark and compared to non-augmented pANO-$n$Z-F12 in Table \ref{tab:W4-17-TAE-RMSDs-results}. As shown in the said table, for atomization energies of neutral molecules, the benefit of adding diffuse functions is negligible for cc-pV$n$Z-F12, but clearly noticeable for the pANO basis sets. (Keep in mind that the most diffuse $s,p$ functions in cc-pV$n$Z-F12 are already decontracted.)

{\scriptsize
\begin{table}[!htbp]
\caption{RMSD (in \kcalmol) for W4-17 benchmark (194$^*$ species) obtained by the new pANO-$n$Z-F12 basis set family at the df-MP2-F12 level.}
\label{tab:W4-17-TAE-RMSDs-results}
\begin{tabular}{|r|l|}
\hline

Basis  & RMSD  \\
\hline \hline

cc-pVDZ-F12          & 1.154 \\
cc-pVTZ-F12          & 0.243 \\
cc-pVQZ-F12          & 0.062 \\
cc-pV5Z-F12       & 0.005 \\
cc-pV5Z-F12rev2      & REF   \\
\hline
aug-cc-pVDZ-F12      & 1.168 \\
aug-cc-pVTZ-F12      & 0.260 \\
aug-cc-pVQZ-F12      & 0.073 \\
\hline
pANO-DZ-econ     & 1.193 \\
pANO-TZ-econ     & 0.518 \\
pANO-QZ-econ     & 0.110 \\
\hline
pANO-DZ-prem     & 1.069 \\
pANO-TZ-prem     & 0.219 \\
pANO-QZ-prem     & 0.087 \\
pANO-5Z          & 0.044 \\
\hline
aug-pANO-DZ-econ & 0.810 \\
aug-pANO-TZ-econ & 0.276 \\
aug-pANO-QZ-econ & 0.074 \\
\hline
aug-pANO-DZ-prem & 0.885 \\
aug-pANO-TZ-prem & 0.178 \\
aug-pANO-QZ-prem & 0.051 \\
aug-pANO-5Z      & 0.019 \\
\hline
\end{tabular}

$*$ Benzene, $S_2$, $S_3$, $S_4$, $P_2$, $P_4$ are excluded.

\end{table}
}

\subsection{Illustrative application: out-of-plane harmonic frequencies of benzene}
It is well known (e.g., \cite{Simandiras1988,jmlm099,jmlm104,Moran2006BenzeneNonplanar}) that linear bending and out-of-plane bending frequencies of unsaturated hydrocarbons are hypersensitive to the basis set. First reported by Handy and coworkers\cite{Simandiras1988}, the phenomenon was ultimately shown to result\cite{jmlm104,Moran2006BenzeneNonplanar} from intramolecular basis set superposition error, and it was also shown\cite{jmlm099,jmlm104} that ANO-type basis sets are more resilient to this type of problem than cc-pV$n$Z or aug-cc-pV$n$Z. (See also McCaslin and Stanton.\cite{McCaslin2013}) 

\begin{figure*}[!h]
     \centering
         \includegraphics[width=0.7\textwidth]{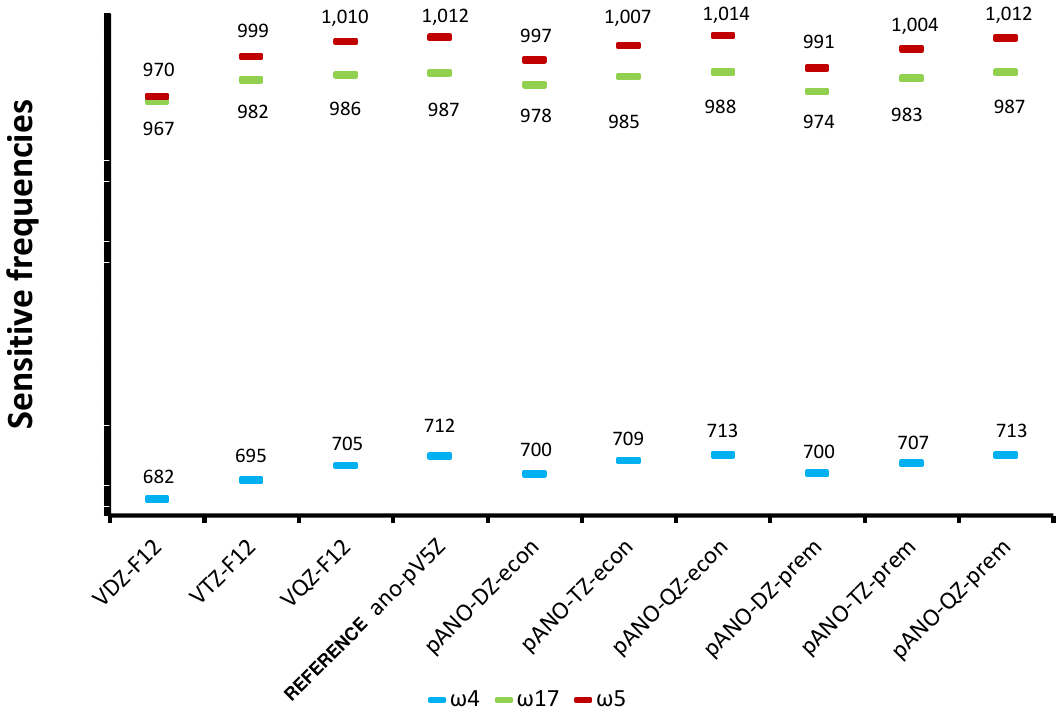}
    \caption{Sensitive out-of-plane frequencies of benzene obtained with different basis sets at the CCSD(T*)(F12*) level.}
    \label{fig:SensFreqsBenzene}
\end{figure*}

Three out-of-plane (OOP) vibrations of benzene (namely $\omega_4$, $\omega_5$, and $\omega_{17}$) are particularly affected.
As shown in Table 5 of Ref. \citenum{jmlm327}, 1 cm$^{-1}$ level basis set convergence at the CCSD(T) level requires ano-pV5Z basis sets,\cite{NeeseValeev2010} although CCSD(T*)(F12*)/cc-pV\{T,Q\}Z-F12 pointwise extrapolation of the potential surface can achieve the same. This is hence a sensitive test for the performance of pANO-$n$Z-F12 basis sets: results for the sensitive OOP modes are summarized in Figure~\ref{fig:SensFreqsBenzene}, while statistics for all harmonic frequencies can be found in Table~\ref{tab:BenzFreqs-RMSDs}.

Clearly, pANO-$n$Z-F12-prem has an edge in performance over the corresponding cc-pV$n$Z-F12 basis sets, where it should be kept in mind that these are the same contracted size. Moreover, the `economy' pANO basis sets appear to outperform the original cc-pV$n$Z-F12. In particular, pANO-DZ-econ and pANO-TZ-econ may be attractive options for F12 frequency calculations on moderately sized molecules.

The RMSD statistics for the benzene frequencies (Table~\ref{tab:BenzFreqs-RMSDs}) show that the new pANO-$n$Z-F12 family provides more accurate results than the original cc-pV$n$Z-F12 in both the \texttt{economy} and \texttt{premium} schemes, at smaller or similar cost, respectively.

\begin{table*}[!htbp]
\centering

\caption{RMSD (in wavenumbers) for benzene harmonic frequencies obtained by pANO-$n$Z-F12 and cc-pV$n$Z-F12 basis sets at the CCSD(T*)(F12*) level relative to CCSD(T)/ano-pV5Z.}
\begin{tabular}{|l|l|l|}
\hline
Basis & RMSD  & RMSD$^*$ \\
\hline
cc-pVDZ-F12 & 12.7 & 11.2  \\
cc-pVTZ-F12 & 5.2 & 4.5  \\
cc-pVQZ-F12 & 1.9 & 1.6  \\
\hline
pANO-DZ-F12-econ & 6.0 & 5.4  \\
pANO-TZ-F12-econ & 2.7 & 2.5  \\
pANO-QZ-F12-econ & 1.7 & 1.7  \\
\hline
pANO-DZ-F12-prem & 7.1 & 6.5  \\
pANO-TZ-F12-prem & 2.9 & 2.6  \\
pANO-QZ-F12-prem & 1.6 & 1.6  \\
\hline
\end{tabular}

$*$ Weighted for degeneracies.

\label{tab:BenzFreqs-RMSDs}
\end{table*}

\subsection{More harmonic frequencies: HFREQ27 and \ce{SO3}}

As a less `narrow' gauge of the performance of pANO-$n$Z-F12 for harmonic frequencies, we evaluated the HFREQ27\cite{jmlm258} benchmark  at the CCSD(T*)(F12*) level of theory. What stands out in Table \ref{tab:HFREQ27-RMSDs} is the general pattern that, as the number of contractions in the pANO-$n$Z-F12 basis increases, the frequencies arrive at the complete basis set limit. The at first startlingly poor RMSD=10 cm$^{-1}$ for pANO-DZ-econ drops to 6 cm$^{-1}$ when the umbrella mode of \ce{NH3} is excised, and further to 5 cm$^{-1}$ if additionally the \ce{HF} and \ce{H2O} stretches are removed; no such drastic effects are seen for other basis sets.

This test shows that pANO-TZ-F12-econ appears to be an attractive low-cost option. It should be kept in mind that valence CCSD(T) at the CBS limit itself has an RMSD over HFREQ27 from the most accurate harmonic frequencies of ca. 5 cm$^{-1}$, owing to neglect of core-valence correlation and post-CCSD(T) correlation effects. Accounting for just the core-valence correlation actually increases RMSD to 7 cm$^{-1}$. It thus follows that if 1-2 cm$^{-1}$ accuracy levels are required, one would anyway have to resort to much more costly electronic structure methods than CCSD(T), and then one might as well upgrade the CCSD(T) treatment without materially adding to the compute budget.

If we add diffuse functions to pANO-DZ-econ, yielding the aug-pANO-DZ-econ basis set, the RMSD is reduced by two-thirds to 3.1 cm$^{-1}$, with no more anomalies left. aug-pANO-DZ-prem has RMSD=5.4 cm$^{-1}$, which however drops to 3.9 cm$^{-1}$ if the lowest bending frequencies of \ce{C2H2} are omitted.

\begin{table*}[!htbp]
\centering
\caption{RMSD (in wavenumbers) for HFREQ27\cite{jmlm258} benchmark frequencies obtained by pANO-$n$Z-F12 and cc-pV$n$Z-F12 basis sets at the CCSD(T*)(F12*) level relative to CCSD(T)/haV\{Q,5\}Z$^a$.}
\begin{tabular}{|l|l|}
\hline\hline
Basis set & RMSD  \\
\hline
cc-pVDZ-F12      & 2.883 \\
cc-pVTZ-F12      & 0.956 \\
cc-pVQZ-F12      & 0.905 \\
\hline

pANO-DZ-econ     & 10.221, 6.354$^b$ \\
pANO-TZ-econ     & 1.976 \\
pANO-QZ-econ     & 1.244 \\
\hline

pANO-DZ-prem     & 4.799 \\
pANO-TZ-prem     & 1.516 \\
pANO-QZ-prem     & 1.237 \\
\hline

aug-pANO-DZ-econ & 3.086 \\
aug-pANO-TZ-econ & 1.079 \\
\hline

aug-pANO-DZ-prem & 5.417 \\
aug-pANO-TZ-prem & 1.414 \\

\hline
valence CCSD(T)/CBS vs. & \\
all-electron FCI/CBS & 4.7\cite{jmlm258} \\
\hline
\hline
\end{tabular}

$^a$ Frequencies were taken from Ref. \citenum{jmlm258} excluding the open-shell SO and \ce{S2} species, as MOLPRO's (F12*) implementation is limited to closed-shell.  \\
$^b$ Excluding the umbrella motion of \ce{NH3}. Additionally removing the stretching frequencies of \ce{HF} and \ce{H2O} further reduces RMSD to 5.094 cm$^{-1}$. 
\label{tab:HFREQ27-RMSDs}
\end{table*}

\begin{table*}
\caption{\ce{SO3} harmonic frequencies at the CCSD(T*)(F12*) level: deviations (cm$^{-1}$) from CCSD(T*)(F12*)/cc-pVQZ-F12 with CABS singles correction, and the effect of the CABS singles correction itself (cm$^{-1}$). Frequencies in spectroscopic order: $\omega_1$ ($a_1'$ stretch), $\omega_2$ ($a_2''$ out-of-plane bend), $\omega_3$ ($e'$ stretch), $\omega_4$ ($e'$ bend).}
\label{tab:SO3}
\begin{tabular}{l|rrrr|rrrr}
\hline
 & \multicolumn{4}{c|}{deviation from reference$^a$} & \multicolumn{4}{c}{CABS singles shift$^b$} \\
Basis set & $\omega_1$ & $\omega_2$ & $\omega_3$ & $\omega_4$ & $\omega_1$ & $\omega_2$ & $\omega_3$ & $\omega_4$ \\
\hline
aug-cc-pVDZ$^c$     & +0.3   & +2.6   & +4.1   & $-$3.9 & +89.8 & +42.5 & +102.9 & +53.9 \\
\quad idem, OptRI CABS$^g$ & $-$14.6 & $-$4.2 & $-$13.3 & $-$7.6 & +74.9 & +35.6 & +85.6 & +50.2 \\
aug-cc-pV(D+d)Z$^d$ & +5.5   & +2.6   & +7.9   & $-$3.3 & +57.6 & +13.6 &  +60.9 & +28.6 \\
cc-pVDZ-F12         & $-$1.4 & $-$1.9 & $-$0.9 & $-$0.2 & +13.6 &  +9.6 &  +14.7 & +11.1 \\
pANO-DZ-econ        & +2.5   & $-$8.6 & +6.0   & +1.2   & +21.7 &  +5.8 &  +29.3 & +10.0 \\
pANO-DZ-prem        & $-$0.9 & $-$6.1 & +0.1   & +0.4   & +16.5 &  +5.1 &  +22.5 & +11.0 \\
\hline
aug-cc-pVTZ         & $-$1.9 & +0.1   & $-$2.2 & $-$0.9 & +26.4 & +17.5 &  +26.5 & +18.8 \\
\quad idem, OptRI CABS$^g$ & $-$2.9 & $-$0.3 & $-$3.5 & $-$1.2 & +25.4 & +17.2 & +25.2 & +18.5 \\
aug-cc-pV(T+d)Z$^d$ & $-$0.6 & +0.1   & $-$0.8 & $-$0.6 & +4.1  & +2.8  &  +2.3  & +5.7  \\
cc-pVTZ-F12         & $-$1.4 & $-$1.0 & $-$1.3 & $-$0.3 &  +4.9 &  +1.2 &   +3.8 &  +1.7 \\
pANO-TZ-econ        & $-$2.1 & +0.4   & $-$1.1 & $-$0.8 &  +3.4 &  +2.2 &   +2.4 &  +2.9 \\
pANO-TZ-prem        & $-$0.3 & $-$0.9 & $-$0.5 & +0.1   &  +2.9 &  +2.0 &   +2.5 &  +2.5 \\
\hline
pANO-QZ-econ        & +0.2   & +0.5   & +0.1   & +0.1   & \multicolumn{4}{c}{---} \\
cc-pVQZ-F12         & \multicolumn{4}{c|}{REF} & \multicolumn{4}{c}{---} \\
\hline
 & \multicolumn{8}{c}{\textbf{conventional CCSD(T)}} \\
 & \multicolumn{4}{c|}{deviation from reference$^a$} & \multicolumn{4}{c}{F12 correlation increment$^e$} \\
aug-cc-pVDZ         & $-$129.9 & $-$53.9 & $-$150.5 & $-$64.9 & +40.4 & +14.1 & +51.6 & +7.0 \\
aug-cc-pV(D+d)Z$^d$ & $-$92.9  & $-$23.9 & $-$103.9 & $-$37.5 & +40.9 & +12.9 & +50.9 & +5.6 \\
aug-cc-pVTZ         & $-$45.5  & $-$22.8 & $-$49.1  & $-$25.1 & +17.2 &  +5.4 & +20.4 & +5.4 \\
aug-cc-pV(T+d)Z$^d$ & $-$21.2  & $-$7.4  & $-$22.2  & $-$11.2 & +16.4 &  +4.7 & +19.1 & +4.9 \\
aug-cc-pVQZ         & $-$22.1  & $-$10.8 & $-$23.7  & $-$10.2 & \multicolumn{4}{c}{---} \\
aug-cc-pV(Q+d)Z$^d$ & $-$9.9   & $-$3.3  & $-$10.7  & $-$3.8  & \multicolumn{4}{c}{---} \\
aug-cc-pV(\{T,Q\}+d)Z$^f$ & $-$1.8 & $-$0.3 & $-$2.5 & +1.5 & \multicolumn{4}{c}{---} \\
\hline\hline
\end{tabular}

\smallskip
\footnotesize
$^a$ Reference values: $\omega_1$=1089.5, $\omega_2$=505.2, $\omega_3$=1423.1, $\omega_4$=534.7~cm$^{-1}$.
$^b$ $\omega$(with CABS singles) $-$ $\omega$(without). Not evaluated for the QZ basis sets, where it is expected to be smaller still than the 1--5~cm$^{-1}$ found for the F12 TZ basis sets.
$^c$ Re-using JKFit basis set as CABS, as per MOLPRO default for the AV$n$Z basis sets
$^d$ aug-cc-pV($n$+d)Z basis set\cite{Wilson2001plusd} on S, aug-cc-pV$n$Z on O.
$^e$ CCSD(T*)(F12*) without CABS singles correction minus conventional CCSD(T), same orbital basis set.
$^f$ $L^{-3}$ extrapolation of the force constants.
$^g$ Using purpose-built aug-cc-pV$n$Z/OptRI basis set\cite{YousafPeterson2009augOptRI} instead of MOLPRO's default for aug-cc-pV$n$Z, namely, aug-cc-pV$n$Z/JKFIT.
All F12 calculations used $\gamma$=1.0. RI basis sets for the CABS: aug-cc-pV$n$Z/JKFIT for aug-cc-pV$n$Z and aug-cc-pV($n$+d)Z (MOLPRO defaults); cc-pV$n$Z-F12/OptRI\cite{YousafPeterson2008OptRI} for both cc-pV$n$Z-F12 and pANO-$n$Z-F12.
\end{table*}

Interestingly enough, the F12 calculations appear to have no specific problem with \ce{SO2}, the geometry and stretching frequencies of which are notoriously sensitive (by up to 35~cm$^{-1}$ at the CCSD(T)/cc-pVTZ level) to high-exponent `inner polarization' $d$ functions.\cite{jmlm107} The role of the latter is\cite{jmlm191} to improve the description of the $3d$ orbital (which in a second-row atom in a high oxidation state lies low enough to act as  an acceptor for back-bonding from oxygen and fluorine lone pairs---an effect that manifests almost entirely at the SCF level). Yet all F12 basis sets considered here, including aug-cc-pVDZ and pANO-DZ-econ, yield CCSD(T*)(F12*) harmonic frequencies within 5~cm$^{-1}$ of one another. We decided to dig a little deeper with a still more pronounced case than \ce{SO2}, namely, the \ce{SO3} molecule.\cite{jmlm121} As one can see in Table \ref{tab:SO3}, CCSD(T*)(F12*) calculations with all basis sets considered -- even simple aug-cc-pVDZ or pANO-DZ-econ -- come within a few cm$^{-1}$ of the best values. (Upon closer inspection, the surprisingly good results for aug-cc-pVDZ stem in part from MOLPRO's decision, possibly for retrocompatibility, to let the aug-cc-pV$n$Z/JKFit Coulomb and exchange fit sets do double-duty as CABS only for aug-cc-pV$n$Z basis sets, rather than using the published aug-cc-pV$n$Z/OptRI sets.\cite{YousafPeterson2009augOptRI} It should be noted that aug-cc-pVDZ/OptRI has a gap in the $d$ exponents for sulfur between 0.9 and 7.2, which is absent in both aug-cc-pVTZ/OptRI and the much denser aug-cc-pVDZ/JKFit sets.) At the standard CCSD(T) level, errors with aug-cc-pVDZ are dismally large, and even the almost 50~cm$^{-1}$ increase in $\omega_3$ from aug-cc-pVDZ to aug-cc-pV(D+d)Z only bridges about one-third of the gap. Aside from the F12 correlation increment (up to ca. 50~cm$^{-1}$ for both basis sets) the CABS singles correction to the SCF energy adds over 100 and 60 cm$^{-1}$, respectively. So the higher d exponents present in the CABS are able to `paper over' the dearth of them in the underlying SCF. Note also that even pANO-DZ-econ only has a 29~cm$^{-1}$ CABS-singles contribution, comparable to that for aug-cc-pVTZ. However, aug-cc-pV(T+d)Z, pANO-TZ-econ, pANO-TZ-prem, and cc-pVTZ-F12 all have small CABS corrections of 1-5 cm$^{-1}$ depending on the mode. pANO-QZ-econ yields nearly identical results to cc-pVQZ-F12.


\section{\label{sec:conclusions}Conclusions}

At the end of this study, we are in a position to state the following:
\begin{enumerate}
    \item We have established that the pseudo-ANO (pANO) approach, in which contraction coefficients are obtained through constrained successive energy minimization, yields basis sets equivalent in quality to traditional ANOs;
    \item We have then leveraged pANO to obtain two sequences of pANO-F12 basis sets, one `economy', the other `premium';
    \item For DZ and TZ, the former is comparable in quality to traditional cc-pV$n$Z-F12 but more compact (and hence more economical), while the latter is equivalent in cost to cc-pV$n$Z-F12 but superior in quality;
    \item F12 basis set convergence is distinctly faster for first-row than for second-row compounds, for the 
    V$n$Z-F12, pANO-$n$Z-F12, and aug-pANO-$n$Z-F12 sequences alike;
    \item A test for the harmonic frequencies of benzene, whose out-of-plane bending modes are plagued by severe intramolecular BSSE, indicates that the pANO-$n$Z-F12 family is more resilient than cc-pV$n$Z-F12;
    \item For the HFREQ27 harmonic frequencies benchmark, CCSD(T*)(F12*)/pANO-DZ-econ performs tolerably well except for the ammonia umbrella mode (which definitely needs more $d$ flexibility). Larger pANO-$n$Z-F12 basis sets perform comparably to cc-pV$n$Z-F12 when viewed in the perspective of a 5 cm$^{-1}$ intrinsic RMSD for the basis set limit (owing to neglect of inner-shell correlation and post-CCSD(T) correlation effects). Interestingly, for pseudohypervalent compounds like \ce{SO2} and \ce{SO3} --- where CCSD(T)/aug-cc-pV$n$Z leads to severe underestimates in stretching frequencies which need to be remedied by adding additional tight (high-exponent) $d$ functions to the basis set --- all basis set families hold their own at the CCSD(T*)(F12*) level, with the CABS singles correction accounting for much of the correction (as long as the complementary auxiliary basis set is adequate);
    \item Augmented pANO-$n$Z-F12 sets significantly outperform their unaugmented pANO-$n$Z-F12 counterparts ($n$=D,T) for noncovalent interactions and provide smaller improvements for total atomization energies; diffuse functions are essential for anion and electron-affinity calculations;
    \item For second-row elements, the aug-cc-pV(5+d)Z-JKFIT Coulomb--exchange fitting basis set is an effective option to suppress the discrepancy with CBS results.
\end{enumerate}

As a final observation, it is clear that the pANO-F12 approach offers the greatest benefits over cc-pV$n$Z-F12 basis sets for double- and triple-zeta.

\begin{acknowledgement}

\noindent This work was supported by the Minerva Foundation, Munich, Germany.  
Vladimir Fishman acknowledges a doctoral fellowship from the Weizmann Institute of Science. JMLM acknowledges the sabbatical hospitality of the Quantum Theory Project, University of Florida, and is grateful for several helpful discussions with its late lamented chair, Prof. John F. Stanton (1961-2025) and with the `Stantonites', particularly Drs. James H. Thorpe, Gregory H. Jones, and Peter R. Franke. 
We also thank Prof. Vincenzo Barone (INSTM, Firenze, Italy) for helpful comments on an earlier draft, which prompted us to contract the $s$ functions as well. We finally thank Dr. Margarita Shepelenko (Martin group) for critically reading the manuscript.
CHEMFARM is supported by the Ben May Center for Chemical Theory and Computation, Weizmann Institute of Science.
\end{acknowledgement}

\begin{suppinfo}

\begin{itemize}
\item Microsoft Excel workbook containing the full raw data.
\item PDF document with contraction schemes for the entire pANO-F12 basis-set family, as well as additional tables and exponents for aug-pANO-F12. 
\item pANO-$n$Z-F12 and aug-pANO-$n$Z-F12 basis-set (.txt) files obtained in this study.
\end{itemize}

\end{suppinfo}

\footnotesize
\bibliography{pANO-F12,f12}

@article{Hylleraas1929,
  author    = {Hylleraas, Egil A.},
  title     = {Neue Berechnung der Energie des Heliums im Grundzustande, sowie des tiefsten Terms von Ortho-Helium},
  journal   = {Zeitschrift f{\"u}r Physik},
  year      = {1929},
  volume    = {54},
  pages     = {347--366},
  doi       = {10.1007/BF01375457},
}

@ARTICLE{Boys1950,
       author = {{Boys}, S.~F.},
        title = "{Electronic Wave Functions. I. A General Method of Calculation for the Stationary States of Any Molecular System}",
      journal = PRS,
         year = 1950,
        month = feb,
       volume = {200},
       number = {1063},
        pages = {542-554},
          doi = {10.1098/rspa.1950.0036},
}

@article{Pekeris1958,
  author    = {Pekeris, C. L.},
  title     = {Ground State of Two-Electron Atoms},
  journal   = PR,
  year      = {1958},
  volume    = {112},
  number    = {5},
  pages     = {1649--1658},
  doi       = {10.1103/PhysRev.112.1649},
}

@article{Pekeris1959,
  author    = {Pekeris, C. L.},
  title     = {$1^{1}S$ and $2^{3}S$ States of Helium},
  journal   = PR,
  year      = {1959},
  volume    = {115},
  number    = {5},
  pages     = {1216--1221},
  doi       = {10.1103/PhysRev.115.1216},
}

@article{Pekeris1962Hminus,
  author    = {Pekeris, C. L.},
  title     = {$1^{1}S$, $2^{1}S$, and $2^{3}S$ States of H$^{-}$ and of He},
  journal   = PR,
  year      = {1962},
  volume    = {126},
  number    = {4},
  pages     = {1470--1476},
  doi       = {10.1103/PhysRev.126.1470},
}

@article{Pekeris1962ExcitedS,
  author    = {Pekeris, C. L.},
  title     = {Excited $S$ States of Helium},
  journal   = PR,
  year      = {1962},
  volume    = {127},
  number    = {2},
  pages     = {509--511},
  doi       = {10.1103/PhysRev.127.509},
}

@article{Jastrow1955,
  title = {Many-Body Problem with Strong Forces},
  author = {Jastrow, Robert},
  journal = PR,
  volume = {98},
  issue = {5},
  pages = {1479--1484},
  numpages = {0},
  year = {1955},
  month = {Jun},
  publisher = {American Physical Society},
  doi = {10.1103/PhysRev.98.1479},
  url = {https://link.aps.org/doi/10.1103/PhysRev.98.1479}
}

@article{BoysHandy1969Indeterminacy,
  author    = {Boys, S. F. and Handy, N. C.},
  title     = {A condition to remove the indeterminacy in interelectronic correlation functions},
  journal   = PRS,
  year      = {1969},
  volume    = {309},
  number    = {1497},
  pages     = {209--220},
  doi       = {10.1098/rspa.1969.0038},
}

@article{BoysHandy1969FullCorr,
  author    = {Boys, S. F. and Handy, N. C.},
  title     = {The determination of energies and wavefunctions with full electronic correlation},
  journal   = PRS,
  year      = {1969},
  volume    = {310},
  number    = {1500},
  pages     = {43--61},
  doi       = {10.1098/rspa.1969.0061},
}

@article{BoysHandy1969Neon,
  author    = {Boys, S. F. and Handy, N. C.},
  title     = {A calculation for the energies and wavefunctions for states of neon with full electronic correlation accuracy},
  journal   = PRS,
  year      = {1969},
  volume    = {310},
  number    = {1500},
  pages     = {63--78},
  doi       = {10.1098/rspa.1969.0062},
}

@article{BoysHandy1969LiH,
  author    = {Boys, S. F. and Handy, N. C.},
  title     = {A first solution, for LiH, of a molecular transcorrelated wave equation by means of restricted numerical integration},
  journal   = PRS,
  year      = {1969},
  volume    = {311},
  number    = {1505},
  pages     = {309--329},
  doi       = {10.1098/rspa.1969.0120},
}

@article{Handy1972TCGaussians,
  author    = {Handy, N. C.},
  title     = {The transcorrelated method for accurate correlation energies using gaussian-type functions: examples on He, H$_2$, LiH and H$_2$O},
  journal   = molphys,
  year      = {1972},
  volume    = {23},
  number    = {1},
  pages     = {1--27},
  doi       = {10.1080/00268977200100011},
}

@article{Alavi2021,
    author = {Schraivogel, Thomas and Cohen, Aron J. and Alavi, Ali and Kats, Daniel},
    title = {Transcorrelated coupled cluster methods},
    journal = jcp,
    volume = {155},
    number = {19},
    pages = {191101},
    year = {2021},
    month = {11},
    issn = {0021-9606},
    doi = {10.1063/5.0072495},
}

@article{Kutzelnigg1985R12,
  author    = {Kutzelnigg, Werner},
  title     = {$r_{12}$-dependent terms in the wave function as closed sums of partial wave amplitudes for large $l$},
  journal   = tca,
  year      = {1985},
  volume    = {68},
  number    = {6},
  pages     = {445--469},
  doi       = {10.1007/BF00527669},
}

@article{KutzelniggKlopper1991R12I,
  author    = {Kutzelnigg, Werner and Klopper, Wim},
  title     = {Wave functions with terms linear in the interelectronic coordinates to take care of the correlation cusp. {I}. General theory},
  journal   = jcp,
  year      = {1991},
  volume    = {94},
  number    = {3},
  pages     = {1985--2001},
  doi       = {10.1063/1.459921},
}

@article{TermathKlopperKutzelnigg1991MP2R12,
  author    = {Termath, Volker and Klopper, Wim and Kutzelnigg, Werner},
  title     = {Wave functions with terms linear in the interelectronic coordinates to take care of the correlation cusp. {II}. Second-order {M}{\o}ller--Plesset (MP2-R12) calculations on closed-shell atoms},
  journal   = jcp,
  year      = {1991},
  volume    = {94},
  number    = {3},
  pages     = {2002--2019},
  doi       = {10.1063/1.459922},
}

@article{NogaKutzelnigg1994CCR12,
  author    = {Noga, Jozef and Kutzelnigg, Werner},
  title     = {Coupled cluster theory that takes care of the correlation cusp by inclusion of linear terms in the interelectronic coordinates},
  journal   = jcp,
  year      = {1994},
  volume    = {101},
  number    = {9},
  pages     = {7738--7762},
  doi       = {10.1063/1.468266},
}

@article{TenNo2004STG,
  author    = {Ten-no, Seiichiro},
  title     = {Initiation of explicitly correlated Slater-type geminal theory},
  journal   = cpl,
  year      = {2004},
  volume    = {398},
  number    = {1-3},
  pages     = {56--61},
  doi       = {10.1016/j.cplett.2004.09.041},
}

@article{Grueneis2017Perspective,
author  = {Gr{\"u}neis, Andreas and Hirata, So and Ohnishi, Yu-ya and Ten-no, Seiichiro},
  title     = {Perspective: Explicitly correlated electronic structure theory for complex systems},
  journal   = jcp,
  year      = {2017},
  volume    = {146},
  pages     = {080901},
  doi       = {10.1063/1.4976974},
}

@article{KutzelniggMorgan1992Rates,
  author  = {Kutzelnigg, Werner and Morgan III, John D.},
  title   = {Rates of convergence of the partial-wave expansions of atomic correlation energies},
  journal = jcp,
  year    = {1992},
  volume  = {96},
  number  = {6},
  pages   = {4484--4508},
  doi     = {10.1063/1.462811},
}

@article{PerssonTaylor1996GTG,
  author  = {Persson, B. Joakim and Taylor, Peter R.},
  title   = {Accurate quantum-chemical calculations: The use of Gaussian-type geminal functions in the treatment of electron correlation},
  journal = jcp,
  year    = {1996},
  volume  = {105},
  number  = {14},
  pages   = {5915--5926},
  doi     = {10.1063/1.472432},
}

@article{Shull1959firstUseOfGeminal,
  author  = {Shull, Harrison},
  title   = {Natural Spin Orbital Analysis of Hydrogen Molecule Wave Functions},
  journal = jcp,
  year    = {1959},
  volume  = {30},
  number  = {6},
  pages   = {1405--1413},
  doi     = {10.1063/1.1730212},
}

@article{Turbomole2020,
  author  = {Balasubramani, Sree Ganesh and Chen, Guo P. and Coriani, Sonia and Diedenhofen, Michael and
             Frank, Marius S. and Franzke, Yannick J. and Furche, Filipp and Grotjahn, Robin and
             Harding, Michael E. and H{\"a}ttig, Christof and Hellweg, Arnim and Helmich-Paris, Benjamin and
             Holzer, Christof and Huniar, Uwe and Kaupp, Martin and Marefat Khah, Alireza and
             Karbalaei Khani, Sarah and M{\"u}ller, Thomas and Mack, Fabian and Nguyen, Brian D. and
             Parker, Shane M. and Perlt, Eva and Rappoport, Dmitrij and Reiter, Kevin and Roy, Saswata and
             R{\"u}ckert, Matthias and Schmitz, Gunnar and Sierka, Marek and Tapavicza, Enrico and
             Tew, David P. and van W{\"u}llen, Christoph and Voora, Vamsee K. and Weigend, Florian and
             Wody{\'n}ski, Artur and Yu, Jason M.},
  title   = {{TURBOMOLE}: Modular program suite for ab initio quantum-chemical and condensed-matter simulations},
  journal = jcp,
  year    = {2020},
  volume  = {152},
  number  = {18},
  pages   = {184107},
  doi     = {10.1063/5.0004635},
}

@article{Molpro2020,
  author  = {Werner, Hans-Joachim and Knowles, Peter J. and Manby, Frederick R. and Black, Joshua A. and Doll, Klaus and He{\ss}elmann, Andreas and Kats, Daniel and K{\"o}hn, Andreas and Korona, Tatiana and Kreplin, David A. and Ma, Qianli and Miller III, Thomas F. and Mitrushchenkov, Alexander and Peterson, Kirk A. and Polyak, Iakov and Rauhut, Guntram and Sibaev, Marat},
  title   = {The {Molpro} quantum chemistry package},
  journal = jcp,
  year    = {2020},
  volume  = {152},
  number  = {14},
  pages   = {144107},
  doi     = {10.1063/5.0005081}
}

@article{Turbomole2023,
  author  = {Franzke, Yannick J. and Holzer, Christof H. and Andersen, Josefine H. and Begu{\v s}i{\'c}, Tomislav and Bruder, Florian and Coriani, Sonia and Della Sala, Fabio and Fabiano, Eduardo and Fedotov, Daniil A. and F{\"u}rst, Susanne and Gillhuber, Sebastian and Grotjahn, Robin and Kaupp, Martin and Kehry, Max and Krsti{\'c}, Marjan and Mack, Fabian and Majumdar, Sourav and Nguyen, Brian D. and Parker, Shane M. and Pauly, Fabian and Pausch, Ansgar and Perlt, Eva and Phun, Gabriel S. and Rajabi, Ahmadreza and Rappoport, Dmitrij and Samal, Bibek and Schrader, Tim and Sharma, Manas and Tapavicza, Enrico and Tre{\ss}, Robert S. and Voora, Vamsee and Wody{\'n}ski, Artur and Yu, Jason M. and Zerulla, Benedikt and Furche, Filipp and H{\"a}ttig, Christof and Sierka, Marek and Tew, David P. and Weigend, Florian},
  title   = {{TURBOMOLE}: Today and Tomorrow},
  journal = jctc,
  year    = {2023},
  volume  = {19},
  number  = {20},
  pages   = {6859--6890},
  doi     = {10.1021/acs.jctc.3c00347}
}

@article{MRCC2020,
  author  = {K{\'a}llay, Mih{\'a}ly and Nagy, P{\'e}ter R. and Mester, D{\'a}vid and Rolik, Zolt{\'a}n and Samu, Gyula and Csontos, J{\'o}zsef and Cs{\'o}ka, J{\'o}zsef and Szab{\'o}, P. Bern{\'a}t and Gyevi-Nagy, L{\'a}szl{\'o} and H{\'e}gely, Bence and Ladj{\'a}nszki, Istv{\'a}n and Szegedy, L{\'o}r{\'a}nt and Lad{\'o}czki, Bence and Petrov, Kl{\'a}ra and Farkas, M{\'a}t{\'e} and Mezei, P{\'a}l D. and Ganyecz, {\'A}d{\'a}m},
  title   = {The {MRCC} program system: Accurate quantum chemistry from water to proteins},
  journal = jcp,
  year    = {2020},
  volume  = {152},
  number  = {7},
  pages   = {074107},
  doi     = {10.1063/1.5142048}
}

@article{MRCC2025,
  author  = {Mester, D{\'a}vid and Nagy, P{\'e}ter R. and Cs{\'o}ka, J{\'o}zsef and Gyevi-Nagy, L{\'a}szl{\'o} and Szab{\'o}, P. Bern{\'a}t and Horv{\'a}th, R{\'e}ka A. and Petrov, Kl{\'a}ra and H{\'e}gely, Bence and Lad{\'o}czki, Bence and Samu, Gyula and L{\H{o}}rincz, Bal{\'a}zs D. and K{\'a}llay, Mih{\'a}ly},
  title   = {Overview of Developments in the {MRCC} Program System},
  journal = jpca,
  year    = {2025},
  volume  = {129},
  number  = {8},
  pages   = {2086--2107},
  doi     = {10.1021/acs.jpca.4c07807}
}

@article{Valeev2004CABS,
  author  = {Valeev, Edward F.},
  title   = {Improving on the resolution of the identity in linear {R12} ab initio theories},
  journal = cpl,
  year    = {2004},
  volume  = {395},
  pages   = {190--195},
  doi     = {10.1016/j.cplett.2004.07.061},
}

@article{HaettigKlopperKoehnTew2012ChemRev,
  author  = {H{\"a}ttig, Christof and Klopper, Wim and K{\"o}hn, Andreas and Tew, David P.},
  title   = {Explicitly Correlated Electrons in Molecules},
  journal = chemrev,
  year    = {2012},
  volume  = {112},
  number  = {1},
  pages   = {4--74},
  doi     = {10.1021/cr200168z},
}

@article{KongBischoffValeev2012ChemRev,
  author  = {Kong, Liguo and Bischoff, Florian A. and Valeev, Edward F.},
  title   = {Explicitly Correlated {R12}/{F12} Methods for Electronic Structure},
  journal = chemrev,
  year    = {2012},
  volume  = {112},
  number  = {1},
  pages   = {75--107},
  doi     = {10.1021/cr200204r},
}

@article{MaWerner2018PNOF12Review,
  author  = {Ma, Qianli and Werner, Hans-Joachim},
  title   = {Explicitly correlated local coupled-cluster methods using pair natural orbitals},
  journal = wires,
  year    = {2018},
  volume  = {8},
  number  = {6},
  pages   = {e1371},
  doi     = {10.1002/wcms.1371},
}

@article{Tew2021PrincipalDomainsF12,
  author    = {Tew, David P.},
  title     = {Principal domains in {F12} explicitly correlated theory},
  journal = aqc,
  year      = {2021},
  volume    = {83},
  pages     = {83--106},
  publisher = {Elsevier},
  doi       = {10.1016/bs.aiq.2021.06.001},
}

@incollection{NagyJensen2017BasisSetsRiCC,
  author    = {Nagy, Bal{\'a}zs and Jensen, Frank},
  title     = {Basis Sets in Quantum Chemistry},
  booktitle = {Reviews in Computational Chemistry},
  editor    = {Parrill, Abby L. and Lipkowitz, Kenny B.},
  volume    = {30},
  pages     = {93--150},
  publisher = {Wiley},
  year      = {2017},
  doi       = {10.1002/9781119356059.ch3},
}

@article{Hill2013GaussianBasisSetsMolecularApplications,
  author  = {Hill, J. Grant},
  title   = {Gaussian Basis Sets for Molecular Applications},
  journal = ijqc,
  year    = {2013},
  volume  = {113},
  number  = {1},
  pages   = {21--34},
  doi     = {10.1002/qua.24355},
}

@article{Jensen2013AtomicOrbitalBasisSets,
  author  = {Jensen, Frank},
  title   = {Atomic orbital basis sets},
  journal = wires,
  year    = {2013},
  volume  = {3},
  number  = {3},
  pages   = {273--295},
  doi     = {10.1002/wcms.1123},
}

@incollection{Peterson2011EIBC_eibc0408,
  title     = {Gaussian Basis Sets for Quantum Mechanical (QM)Calculations},
  author = {Peterson, Kirk A.},
  booktitle = {Encyclopedia of Inorganic and Bioinorganic Chemistry},
  editor    = {Scott, Robert A.},
  publisher = {John Wiley \& Sons, Ltd.},
  year      = {2011},
  doi       = {10.1002/9781119951438.eibc0408},
}

@article{jmlm269,
author = {Sylvetsky, Nitai and Peterson, Kirk A. and Karton, Amir and Martin, Jan M. L.},
doi = {10.1063/1.4952410},
issn = {0021-9606},
journal = jcp,
month = {jun},
number = {21},
pages = {214101},
title = {{Toward a W4-F12 approach: Can explicitly correlated and orbital-based ab initio CCSD(T) limits be reconciled?}},
url = {http://arxiv.org/abs/1605.03398 http://scitation.aip.org/content/aip/journal/jcp/144/21/10.1063/1.4952410 http://aip.scitation.org/doi/10.1063/1.4952410},
volume = {144},
year = {2016}
}

@article{KohnTew2010CuspI,
  author  = {K{\"o}hn, Andreas and Tew, David P.},
  title   = {Explicitly correlated coupled-cluster theory using cusp conditions. {I}. {P}erturbation analysis of coupled-cluster singles and doubles ({CCSD-F12})},
  journal = jcp,
  year    = {2010},
  volume  = {133},
  number  = {17},
  pages   = {174117},
  doi     = {10.1063/1.3496372}
}

@article{jmlm325,
   shorthand={325},
  title = {Correlation Consistent Basis Sets for Explicitly Correlated Theory: The Transition Metals},
  volume = {19},
  ISSN = {1549-9626},
  DOI = {10.1021/acs.jctc.3c00506},
  number = {17},
  journal = jctc,
  publisher = {American Chemical Society (ACS)},
  author = {Semidalas,  Emmanouil and Martin,  Jan M. L.},
  year = {2023},
  pages = {5806-5820}
}

@article{jmlm312,
   shorthand = {312},
   author = {Nisha Mehta and Jan M. L. Martin},
   doi = {10.1021/acs.jpca.2c02494},
   issn = {1089-5639},
   issue = {24},
   journal = jpca,
   month = {6},
   pages = {3964-3971},
   title = {MP2-F12 Basis Set Convergence near the Complete Basis Set Limit: Are h Functions Sufficient?},
   volume = {126},
   year = {2022},
}

@article{jmlm314,
   shorthand = {314},
   author = {Nisha Mehta and Jan M L Martin},
   doi = {10.1021/acs.jctc.2c00426},
   issn = {1549-9618},
   issue = {10},
   journal = jctc,
   month = {10},
   pages = {5978-5991},
   title = {Explicitly Correlated Double-Hybrid DFT: A Comprehensive Analysis of the Basis Set Convergence on the GMTKN55 Database},
   volume = {18},
   year = {2022},
}

@article{jmlm318,
   shorthand = {318},
   author = {Nisha Mehta and Jan M. L. Martin},
   doi = {10.1021/acs.jpclett.2c02620},
   issn = {1948-7185},
   issue = {40},
   journal = jpcl,
   month = {10},
   pages = {9332-9338},
   title = {Reduced-Scaling Double Hybrid Density Functional Theory with Rapid Basis Set Convergence through Localized Pair Natural Orbital F12},
   volume = {13},
   year = {2022},
}

@article{jmlm290,
   shorthand = {290},
   author = {Jan M L Martin and Golokesh Santra},
   doi = {10.1002/ijch.201900114},
   issn = {0021-2148},
   issue = {8-9},
   journal = ijc,
   month = {12},
   pages = {787-804},
   title = {Empirical Double‐Hybrid Density Functional Theory: A ‘Third Way’ in Between WFT and DFT},
   volume = {60},
   year = {2020},
}

@article{jmlm261,
    shorthand={261},
   author = {Kirk A. Peterson and Manoj K. Kesharwani and Jan M. L. Martin},
   doi = {10.1080/00268976.2014.985755},
   issn = {0026-8976},
   issue = {13-14},
   journal = molphys,
   month = {7},
   pages = {1551-1558},
   title = {The cc-pV5Z-F12 basis set: reaching the basis set limit in explicitly correlated calculations},
   volume = {113},
   year = {2015},
}

@article{jmlm099,
   author = {Jan M. L. Martin and Peter R. Taylor and Timothy J. Lee},
   doi = {10.1016/S0009-2614(97)00735-5},
   issn = {00092614},
   issue = {3-4},
   journal = cpl,
   shorthand = {099},
   month = {8},
   pages = {414-422},
   title = {The harmonic frequencies of benzene. A case for atomic natural orbital basis sets},
   volume = {275},
   year = {1997},
}

@article{jmlm104,
   shorthand={104},
   author = {Jan M. L. Martin and Timothy J. Lee and Peter R. Taylor},
   doi = {10.1063/1.475429},
   issn = {0021-9606},
   issue = {2},
   journal = jcp,
   month = {1},
   pages = {676-691},
   title = {A purely ab initio spectroscopic quality quartic force field for acetylene},
   volume = {108},
   year = {1998},
}

@article{DolgCao2012RelativisticPseudopotentials,
  author  = {Dolg, Michael and Cao, Xiaoyan},
  title   = {Relativistic {Pseudopotentials}: Their Development and Scope of Applications},
  journal = chemrev,
  year    = {2012},
  volume  = {112},
  number  = {1},
  pages   = {403--480},
  month   = jan,
  doi     = {10.1021/cr2001383},
  url     = {https://doi.org/10.1021/cr2001383},
}

@article{Hill2014,
  author  = {Hill, J. G. and Peterson, K. A.},
  title   = {{Correlation consistent basis sets for explicitly correlated wavefunctions: Pseudopotential-based basis sets for the post-d main group elements Ga–Rn}},
  journal = jcp,
  volume  = {141},
  pages   = {094106},
  year    = {2014},
  doi     = {10.1063/1.4893989}
}

@article{Hill2021,
  author  = {Hill, J. G. and Shaw, R. A.},
  title   = {{Correlation consistent basis sets for explicitly correlated wavefunctions: Pseudopotential-based basis sets for the group 11 (Cu, Ag, Au) and 12 (Zn, Cd, Hg) elements}},
  journal = jcp,
  volume  = {155},
  pages   = {174113},
  year    = {2021},
  doi     = {10.1063/5.0070638}
}

@article{McCaslin2013,
  author  = {McCaslin, Laura and Stanton, John F.},
  title   = {Calculation of fundamental frequencies for small polyatomic molecules: a comparison between correlation consistent and atomic natural orbital basis sets},
  journal = molphys,
  volume  = {111},
  pages   = {1492--1496},
  year    = {2013},
  doi     = {10.1080/00268976.2013.811303}
}

@article{NeeseValeev2010,
  author  = {Neese, Frank and Valeev, Edward F.},
  title   = {Revisiting the atomic natural orbital approach for basis sets},
  journal = jctc,
  volume  = {7},
  pages   = {33--43},
  year    = {2011},
  doi     = {10.1021/ct100396y}
}

@article{Persson1997P4,
  author  = {Persson, B. Joakim and Taylor, Peter R. and Lee, Timothy J.},
  title   = {Ab initio geometry, quartic force field, and vibrational frequencies for P$_4$},
  journal = jcp,
  year    = {1997},
  volume  = {107},
  number  = {13},
  pages   = {5051--5057},
  doi     = {10.1063/1.474868}
}

@article{Kohn2010CCSDF12star,
  author  = {H{\"a}ttig, Christof and Tew, David P. and K{\"o}hn, Andreas},
  title   = {Accurate and efficient approximations to explicitly correlated coupled-cluster singles and doubles theory},
  journal = jcp,
  year    = {2010},
  volume  = {132},
  number  = {23},
  pages   = {231102},
  doi     = {10.1063/1.3442368}
}

@article{jmlm285,
   shorthand={285},
   author = {Manoj K. Kesharwani and Nitai Sylvetsky and Andreas Köhn and David P. Tew and Jan M. L. Martin},
   doi = {10.1063/1.5048665},
   issn = {0021-9606},
   issue = {15},
   journal = jcp,
   month = {10},
   pages = {154109},
   title = {Do CCSD and approximate CCSD-F12 variants converge to the same basis set limits? The case of atomization energies},
   volume = {149},
   year = {2018},
}

@article{jmlm272,
    shorthand={272},
   author = {Debashree Manna and Manoj Kumar Kesharwani and Nitai Sylvetsky and Jan M. L. Martin},
   doi = {10.1021/acs.jctc.6b01046},
   issn = {1549-9618},
   issue = {7},
   journal = jctc,
   month = {7},
   pages = {3136-3152},
   title = {Conventional and Explicitly Correlated ab Initio Benchmark Study on Water Clusters: Revision of the BEGDB and WATER27 Data Sets},
   volume = {13},
   year = {2017},
}

@article{jmlm276,
  title={The aug-cc-pVnZ-F12 basis set family: Correlation consistent basis sets for explicitly correlated benchmark calculations on anions and noncovalent complexes},
  author = {Nitai Sylvetsky and Manoj K. Kesharwani and  Jan M. L. Martin},
  journal=jcp,
  volume={147},
  number={13},
  pages = {134106},
  year={2017},
  doi={10.1063/1.4998332},
  publisher={AIP Publishing},
}

@article{Hill2009_extrap_MP2_CCSD_F12,
  author  = {Hill, J. Grant and Peterson, Kirk A. and Knizia, Gerald and Werner, Hans-Joachim},
  title   = {Extrapolating MP2 and CCSD explicitly correlated correlation energies to the complete basis set limit with first and second row correlation consistent basis sets},
  journal = jcp,
  year    = {2009},
  volume  = {131},
  pages   = {194105},
  doi     = {10.1063/1.3265857}
}

@article{Weigend2002_RIHF_direct,
  author  = {Weigend, Florian},
  title   = {A fully direct RI-HF algorithm: Implementation, optimised auxiliary basis sets, demonstration of accuracy and efficiency},
  journal = pccp,
  year    = {2002},
  volume  = {4},
  pages   = {4285--4291},
  doi     = {10.1039/b204199p}
}

@article{Weigend2006_coulomb_fitting,
  author  = {Weigend, Florian},
  title   = {Accurate Coulomb-fitting basis sets for H to Rn},
  journal = pccp,
  doi = {10.1039/B515623H},
  year    = {2006},
  volume  = {8},
  pages   = {1057--1065}
}

@article{Nash2023_ccpVndZ_JKFit,
  author  = {Nash, Harry W. and Shaw, Robert A. and Hill, J. Grant},
  title   = {Correlation consistent auxiliary basis sets in density fitting Hartree--Fock: The atoms sodium through argon revisited},
  journal = jcc,
  year    = {2023},
  volume  = {44},
  number  = {11},
  pages   = {1119--1128},
  doi     = {10.1002/jcc.27069}
}

@article{ShawHill2017CABSsingles,
  author  = {Shaw, Robert A. and Hill, J. Grant},
  title   = {Approaching the {Hartree}--{Fock} Limit through the Complementary Auxiliary Basis Set Singles Correction and Auxiliary Basis Sets},
  journal = jctc,
  year    = {2017},
  volume  = {13},
  number  = {4},
  pages   = {1691--1698},
  doi     = {10.1021/acs.jctc.7b00140}
}

@string{aqc = "Adv. Quantum Chem."}

@string{ijc = "Isr. J. Chem."}

@string{cpl = "Chem. Phys. Lett."}

@string{jcp = "J. Chem. Phys."}

@string{jcc = "J. Comput. Chem."}

@string{pccp = "Phys. Chem. Chem. Phys."}

@string{PCCP = "Phys. Chem. Chem. Phys."}

@string{jpca = "J. Phys. Chem. A"}

@string{jpcl = "J. Phys. Chem. Lett."}

@string{tca = "Theor. Chim. Acta"}

@string{TCA = "Theor. Chem. Acc."}

@string{ijqc = "Int. J. Quantum Chem."}

@string{theochem = "J. Mol. Struct. {\sc (Theochem)}"}

@string{jacs= "J. Am. Chem. Soc."}

@string{JACS= "J. Am. Chem. Soc."}

@string{PR= "Phys. Rev."}

@string{PRS= "Proc. R. Soc. London A"}

@string{jctc= "J. Chem. Theory Comput."}

@string{IJQC="Int. J. Quantum Chem."}

@string{molphys= "Mol. Phys."}

@string{chemrev= "Chem. Rev."}

@string{wires= "Wiley Interdiscip. Rev.: Comput. Mol. Sci."}

@article{jmlm340,
      title={Coupling between thermochemical contributions of subvalence correlation and of higher-order post-CCSD(T) correlation effects --- a step toward `W5 theory'},
author={Aditya Barman and Gregory H. Jones and Kaila E. Weflen and Margarita Shepelenko and Jan M. L. Martin},
      year={2026},
      journal = jpca,
      volume = {130},
      pages = {2943-2955},
      doi = {10.1021/acs.jpca.6c00467},
}

@article{Simandiras1988,
  author  = {Simandiras, Emmanuel D. and Rice, Julia E. and Lee, Timothy J. and Amos, Roger D. and Handy, Nicholas C.},
  title   = {On the necessity of f basis functions for bending frequencies},
  journal = jcp,
  volume  = {88},
  pages   = {3187--3195},
  year    = {1988},
  doi     = {10.1063/1.453963},
}

@article{Moran2006BenzeneNonplanar,
  author  = {Moran, Damian and Simmonett, Andrew C. and Leach, Franklin E. and Allen, Wesley D. and von Ragu\'e Schleyer, Paul and Schaefer, Henry F.},
  title   = {Popular Theoretical Methods Predict Benzene and Arenes To Be Nonplanar},
  journal = jacs,
  year    = {2006},
  volume  = {128},
  number  = {29},
  pages   = {9342--9343},
  doi     = {10.1021/ja0630285}
}

@article{jmlm327,
   shorthand={327},
   author = {Emmanouil Semidalas and Jan M. L. Martin},
   doi = {10.1080/00268976.2023.2263593},
   journal = molphys,
   month = {8},
   pages = {e2263593},
   title = {Can G4-like Composite Ab Initio Methods Accurately Predict Vibrational Harmonic Frequencies?},
   volume = {122},
   year = {2024},
   note = {[Timothy J. Lee memorial issue]},
}

@article{peterson2008systematically,
  title   = {Systematically convergent basis sets for explicitly correlated wavefunctions: The atoms H, He, B--Ne, and Al--Ar},
  author  = {Peterson, Kirk A and Adler, Thomas B and Werner, Hans-Joachim},
  journal = jcp,
  volume  = {128},
  number  = {8},
  pages   = {084102},
  year    = {2008},
  doi     = {10.1063/1.2831537},
  publisher = {AIP Publishing}
}

@article{dunning1989gaussian,
  title   = {Gaussian basis sets for use in correlated molecular calculations. I. The atoms boron through neon and hydrogen},
  author  = {Dunning Jr, Thom H},
  journal = jcp,
  volume  = {90},
  number  = {2},
  pages   = {1007--1023},
  year    = {1989},
  doi     = {10.1063/1.456153},
  publisher = {American Institute of Physics}
}

@article{dunning1996gaussian6,
title = {Gaussian basis sets for use in correlated molecular calculations. VI. Sextuple zeta correlation consistent basis sets for boron through neon},
journal = theochem,
volume = {388},
pages = {339-349},
year = {1996},
issn = {0166-1280},
doi = {10.1016/S0166-1280(96)80048-0},
url = {https://www.sciencedirect.com/science/article/pii/S0166128096800480},
author = {Wilson, Angela K and van Mourik, Tanja and Dunning Jr, Thom H}
}

@article{dunning2000gaussian7,
author = {Van Mourik, Tanja and Dunning Jr., Thom H.},
title = {Gaussian basis sets for use in correlated molecular calculations. VIII. Standard and augmented sextuple zeta correlation consistent basis sets for aluminum through argon},
journal = ijqc,
volume = {76},
number = {2},
pages = {205-221},
year = {2000},
doi = {10.1002/(SICI)1097-461X(2000)76:2<205::AID-QUA10>3.0.CO;2-C},
}

@article{almlof1987general,
  title   = {General contraction of Gaussian basis sets. I. Atomic natural orbitals for first- and second-row atoms},
  author  = {Alml{\"o}f, Jan and Taylor, Peter R},
  journal = jcp,
  volume  = {86},
  number  = {7},
  pages   = {4070--4077},
  year    = {1987},
  doi     = {10.1063/1.452775}
}

@article{AlmlofTaylorPart2,
  author  = {Alml{\"o}f, Jan and Taylor, Peter R.},
  title   = {General contraction of Gaussian basis sets. II. Atomic natural orbitals and the calculation of atomic and molecular properties},
  journal = jcp,
  volume  = {92},
  pages   = {551--560},
  year    = {1990},
  doi     = {10.1063/1.458458},
}

@article{klopper2002explicitly,
  title   = {Explicitly correlated second-order M{\o}ller--Plesset methods with auxiliary basis sets},
  author  = {Klopper, Wim and Samson, Claire C. M.},
  journal = jcp,
  volume  = {116},
  number  = {15},
  pages   = {6397--6410},
  year    = {2002},
  doi     = {10.1063/1.1467341},
  publisher = {American Institute of Physics}
}

@article{ranasinghe2013ccsd,
  title   = {CCSD(T)/CBS atomic and molecular benchmarks for H through Ar},
  author  = {Ranasinghe, Duminda S and Petersson, George A},
  journal = jcp,
  volume  = {138},
  number  = {14},
  pages   = {144104},
  year    = {2013},
  doi     = {10.1063/1.4798707},
  publisher = {AIP Publishing}
}

@article{nelder1965simplex,
  title   = {A simplex method for function minimization},
  author  = {Nelder, John A and Mead, Roger},
  journal = {The Computer Journal},
  volume  = {7},
  number  = {4},
  pages   = {308--313},
  year    = {1965},
  doi     = {10.1093/comjnl/7.4.308},
  publisher = {The British Computer Society}
}

@article{jmlm273,
  title   = {W4-17: A diverse and high-confidence dataset of atomization energies for benchmarking high-level electronic structure methods},
  author  = {Karton, Amir and Sylvetsky, Nitai and Martin, Jan M.~L.},
  journal = jcc,
  volume  = {38},
  number  = {24},
  pages   = {2063--2075},
  year    = {2017},
  doi     = {10.1002/jcc.24854},
  publisher = {Wiley Online Library}
}

@article{rezac2011s66,
  title   = {S66: A well-balanced database of benchmark interaction energies relevant to biomolecular structures},
  author  = {Rez{\'a}c, Jan and Riley, Kevin E. and Hobza, Pavel},
  journal = jctc,
  volume  = {7},
  number  = {8},
  pages   = {2427--2438},
  year    = {2011},
  doi     = {10.1021/ct2002946},
  publisher = {ACS Publications}
}

@article{Widmark1990,
    title = {{Density matrix averaged atomic natural orbital (ANO) basis sets for correlated molecular wave functions - I. First row atoms}},
    year = {1990},
    journal = tca,
    author = {Widmark, Per Olof and Malmqvist, Per {\AA}ke and Roos, Bj\"orn O.},
    number = {5},
    pages = {291--306},
    volume = {77},
    isbn = {0040-5744},
    doi = {10.1007/BF01120130},
    issn = {00405744}
}

@article{Kendall1992,
    title = {{Electron affinities of the first-row atoms revisited. Systematic basis sets and wave functions}},
    year = {1992},
    journal = jcp,
    author = {Kendall, Rick A. and Dunning, Thom H. and Harrison, Robert J.},
    number = {9},
    pages = {6796--6806},
    volume = {96},
    url = {http://scitation.aip.org/content/aip/journal/jcp/96/9/10.1063/1.462569},
    doi = {10.1063/1.462569},
    issn = {00219606}
}

@article{Marchetti2009,
    title = {{Accurate Calculations of Intermolecular Interaction Energies Using Explicitly Correlated Coupled Cluster Wave Functions and a Dispersion-Weighted MP2 Method}},
    journal = jpca,
    year = {2009},
    author = {Marchetti, Oliver and Werner, Hans-Joachim},
    pages = {11580--11585},
    volume = {113},
    doi={10.1021/jp9059467},
}

@article{jmlm258,
    title = {{Assessment
of CCSD(T)-F12 Approximations and Basis
Sets for Harmonic Vibrational Frequencies}},
    journal = jctc,
    year = {2014},
    author = {Martin, Jan M. L. and Kesharwani, Manoj K.},
    pages = {2085-2090},
    volume = {10},
    number = {5},
    issn = {1549-9618},
    doi = {10.1021/ct500174q},
}

@article{Valeev2025slimmer,
  author  = {Powell, Samuel R. and Surjuse, Kshitijkumar A. and Gaudel, Bimal and Valeev, Edward F.},
  title   = {Slimmer Geminals For Accurate {F12} Electronic Structure Models},
  journal = jctc,
  year    = {2025},
  volume  = {21},
  pages   = {8833--8842},
  doi     = {10.1021/acs.jctc.5c00971}
}

@article{Frisch1986HBond,
  author  = {Frisch, Michael J. and Del Bene, Janet E. and Binkley, J. Stephen and Schaefer, III, Henry F.},
  title   = {Extensive theoretical studies of the hydrogen-bonded complexes {(H$_2$O)$_2$}, {(H$_2$O)$_2$H$^+$}, {(HF)$_2$}, {(HF)$_2$H$^+$}, {F$_2$H$^-$}, and {(NH$_3$)$_2$}},
  journal = jcp,
  year    = {1986},
  volume  = {84},
  pages   = {2279--2289},
    doi     = {10.1063/1.450390}
}

@article{jmlm191,
    title = {{Heats of formation of perchloric acid, \ce{HClO4}, and perchloric anhydride, \ce{Cl2O7}. Probing the limits of W1 and W2 theory}},
    year = {2006},
    journal = theochem,
    author = {Martin, Jan M. L.},
    number = {1-3},
    month = {10},
    pages = {19--26},
    volume = {771}, 
    doi = {10.1016/j.theochem.2006.03.035},
    issn = {01661280}
}

@article{jmlm107,
    title = {{Basis set convergence study of the atomization energy, geometry, and anharmonic force field of SO2: The importance of inner polarization functions}},
    year = {1998},
    journal = jcp,
    author = {Martin, Jan M. L.},
    number = {7},
    month = {2},
    pages = {2791--2800},
    volume = {108},
    doi = {10.1063/1.475670},
    issn = {0021-9606}
}

@article{YousafPeterson2008OptRI,
  author  = {Yousaf, Kazim E. and Peterson, Kirk A.},
  title   = {Optimized auxiliary basis sets for explicitly correlated methods},
  journal = jcp,
  year    = {2008},
  volume  = {129},
  number  = {18},
  pages   = {184108},
  doi     = {10.1063/1.3009271}
}

@article{YousafPeterson2009augOptRI,
  author  = {Yousaf, Kazim E. and Peterson, Kirk A.},
  title   = {Optimized complementary auxiliary basis sets for explicitly correlated methods: aug-cc-p{V}n{Z} orbital basis sets},
  journal = cpl,
  year    = {2009},
  volume  = {476},
  number  = {4--6},
  pages   = {303--307},
  doi     = {10.1016/j.cplett.2009.06.003}
}

@article{Wilson2001plusd,
  title = {Gaussian basis sets for use in correlated molecular calculations. X. The atoms aluminum through argon revisited},
  author = {Dunning, Thom H. and Peterson, Kirk A. and Wilson, Angela K.},
  journal = jcp,
  year = {2001},
  month = jun,
  volume = {114},
  number = {21},
  pages = {9244-9253},
  publisher = {AIP Publishing},
  doi = {10.1063/1.1367373}
}

@article{jmlm121,
   shorthand={121},
   author = {Jan M. L. Martin},
   doi = {10.1016/S1386-1425(98)00271-6},
   issn = {13861425},
   issue = {3},
   journal = {Spectrochimica Acta Part A: Molecular and Biomolecular Spectroscopy},
   month = {3},
   pages = {709-718},
   title = {A fully ab initio quartic force field of spectroscopic quality for SO3},
   volume = {55},
   year = {1999},
}

@article{Cattell1966scree,
  author  = {Cattell, Raymond B.},
  title   = {The Scree Test For The Number Of Factors},
  journal = {Multivariate Behavioral Research},
  volume  = {1},
  number  = {2},
  pages   = {245--276},
  year    = {1966},
  doi     = {10.1207/s15327906mbr0102_10}
}

\end{document}


\noindent\textbf{Contents}
\begin{itemize}
\item Tables S1--S2: diffuse-function exponents for the aug-pANO-$n$Z-F12 basis sets.
\item Tables S3--S12: conventional ANO contractions of the 7ZaPa primitive set for the angular-correlation ($d$--$i$) functions of B--F and Al--Cl, underlying Tables 1 and 2 of the main text.
\item Tables S13--S22: the corresponding pseudo-ANO (pANO) contractions obtained by constrained energy minimization.
\item Table S23: RMSDs for an alternative hydrogen contraction scheme derived from cc-pV5Z-F12.
\end{itemize}
The pANO-$n$Z-F12 basis sets themselves are provided in MOLPRO format as separate .txt files.

\section{Exponents for aug-pANO-$n$Z-F12}

{\scriptsize
\begin{table}[H]
\caption{Optimized exponents ($\zeta$) for the orbital functions in the aug-pANO-$n$Z-F12 basis sets, $n$=D,T.}
\begin{tabular}{|l|l|l|l|l|l|l|l|}
\hline
\multicolumn{6}{|c|}{aug-pANO-DZ-econ-F12} \\
\hline
H & B & C & N & O & F \\
p,0.121502 & d,0.128257 & d,0.170060 & d,0.249901 & d,0.367735 & d,0.523518 \\

\hline

  & Al & Si & P & S & Cl \\
 & d,0.059946 & d,0.094790 & d,0.130571 & d,0.162674 & d,0.207169 \\
\hline

\multicolumn{6}{|c|}{aug-pANO-DZ-prem-F12} \\
\hline
H  & B  & C  & N  & O  & F \\
p,0.105921 & d,0.097129 & d,0.141424 & d,0.160741 & d,0.216615 & d,0.370581 \\

\hline

  & Al & Si & P & S & Cl \\
 & d,0.045516 & d,0.096554 & d,0.107051 & d,0.142261 & d,0.179195 \\

\hline

\multicolumn{6}{|c|}{aug-pANO-TZ-econ-F12} \\
\hline
H & B & C & N & O & F \\
p,0.109701 & d,0.100491 & d,0.142348 & d,0.166967 & d,0.216085 & d,0.331438 \\
d,0.145102 & f,0.153757 & f,0.214259 & f,0.353926 & f,0.514409 & f,0.637358 \\

\hline

  & Al & Si & P & S & Cl \\
  & d,0.046421 & d,0.067295 & d,0.120143 & d,0.149940 & d,0.153117 \\
  & f,0.085613 & f,0.127616 & f,0.122185 & f,0.158364 & f,0.192896 \\
\hline

\multicolumn{6}{|c|}{aug-pANO-TZ-prem-F12} \\
\hline
H & B & C & N & O & F \\
p,0.111232 & d,0.086450 & d,0.131893 & d,0.154145 & d,0.176739 & d,0.223211 \\
d,0.137542 & f,0.121490 & f,0.115534 & f,0.147782 & f,0.201999 & f,0.261950 \\

\hline

  & Al & Si & P & S & Cl \\
  & d,0.042728 & d,0.054516 & d,0.120548 & d,0.079527 & d,0.184178 \\
  & f,0.069716 & f,0.094338 & f,0.127924 & f,0.153038 & f,0.204012 \\
\hline

\end{tabular}

\end{table}
}

{\scriptsize
\begin{table}[H]
\caption{Optimized exponents ($\zeta$) for the orbital functions in the aug-pANO-$n$Z-F12 basis sets, $n$=Q,5.}
\begin{tabular}{|l|l|l|l|l|l|l|l|}
\hline
\multicolumn{6}{|c|}{aug-pANO-QZ-econ-F12} \\
\hline
H & B & C & N & O & F \\
p,0.110811 & d,0.085058 & d,0.132316 & d,0.163899 & d,0.191067 & d,0.225075 \\
d,0.123123 & f,0.124440 & f,0.156976 & f,0.198425 & f,0.270302 & f,0.394121 \\
f,0.154988 & g,0.166285 & g,0.390916 & g,0.406242 & g,0.390977 & g,0.523835 \\

\hline

  & Al & Si & P & S & Cl \\
  & d,0.042906 & d,0.056789 & d,0.120695 & d,0.072348 & d,0.092943 \\
  & f,0.070157 & f,0.095697 & f,0.127382 & f,0.153339 & f,0.190992 \\
  & g,0.102803 & g,0.162923 & g,0.191394 & g,0.207338 & g,0.257638 \\
\hline

\multicolumn{6}{|c|}{aug-pANO-QZ-prem-F12} \\
\hline
H & B & C & N & O & F \\
p,0.103606 & d,0.064256 & d,0.076582 & d,0.087787 & d,0.108453 & d,0.149157 \\
d,0.111472 & f,0.106602 & f,0.124205 & f,0.120113 & f,0.188105 & f,0.421965 \\
f,0.146308 & g,0.118760 & g,0.114330 & g,0.157103 & g,0.265786 & g,0.459087 \\

\hline

  & Al & Si & P & S & Cl \\
  & d,0.043565 & d,0.095949 & d,0.117095 & d,0.068915 & d,0.088782 \\
  & f,0.046399 & f,0.059872 & f,0.079597 & f,0.090684 & f,0.116826 \\
  & g,0.087078 & g,0.096803 & g,0.131486 & g,0.180128 & g,0.255707 \\
\hline

\multicolumn{6}{|c|}{aug-pANO-5Z-F12} \\
\hline
H & B & C & N & O & F \\
p,0.106129 & d,0.064809 & d,0.076836 & d,0.088186 & d,0.109573 & d,0.151712 \\
d,0.113028 & f,0.111223 & f,0.127476 & f,0.152059 & f,0.192701 & f,0.257130 \\
f,0.180857 & g,0.154354 & g,0.195582 & g,0.230273 & g,0.298558 & g,0.440567 \\
g,0.240363 & h,0.182370 & h,0.158094 & h,0.208550 & h,0.278434 & h,0.312444 \\

\hline

  & Al & Si & P & S & Cl \\
  & d,0.042386 & d,0.094755 & d,0.116176 & d,0.073412 & d,0.090557 \\
  & f,0.049167 & f,0.064134 & f,0.085667 & f,0.098026 & f,0.126489 \\
  & g,0.087580 & g,0.127265 & g,0.144012 & g,0.181253 & g,0.233617 \\
  & h,0.169619 & h,0.234541 & h,0.191202 & h,0.249636 & h,0.335144 \\
\hline

\end{tabular}

\end{table}
}

\newpage
\section{ANOs for the angular-correlation ($d$--$i$) functions of B--F and Al--Cl}
 The 7ZaPa primitive exponents (Ranasinghe and Petersson, J. Chem. Phys. \textbf{138}, 144104 (2013)) are shown rounded to three significant figures; full-precision values are available from the Basis Set Exchange.

{\scriptsize
\begin{table}[H]
\caption{Conventional ANO contractions of the 7ZaPa primitive set for boron}
\resizebox{\textwidth}{!}{\begin{tabular}{|l|l|rrrrrrr||l|}
\hline\hline

d & B   & {\color[HTML]{00B050} 3.64E+00} & {\color[HTML]{00B050} 1.84E+00} & {\color[HTML]{00B050} 9.33E-01} & {\color[HTML]{00B050} 4.72E-01} & {\color[HTML]{00B050} 2.39E-01} & {\color[HTML]{00B050} 1.21E-01} & {\color[HTML]{00B050} 6.13E-02} & NO occ                                 \\
c & 1.7 & 0.007859                        & 0.016698                        & 0.124297                        & 0.338237                        & 0.418031                        & 0.223248                        & 0.046856                        & {\color[HTML]{00B0F0} 0.0041737}       \\
c & 1.7 & -0.005343                       & -0.041749                       & 0.542165                        & 0.608952                        & -0.480692                       & -0.470299                       & -0.120616                       & {\color[HTML]{00B0F0} 0.000059}        \\
c & 1.7 & 0.01883                         & 0.294686                        & 1.090826                        & -1.213753                       & -0.378803                       & 0.649318                        & 0.213912                        & {\color[HTML]{00B0F0} 0.0000103}       \\
c & 1.7 & 0.028559                        & 1.468213                        & -1.280665                       & -0.779141                       & 2.00353                         & -1.091223                       & -0.16653                        & {\color[HTML]{00B0F0} 0.0000013}       \\
c & 1.7 & 0.885839                        & 0.314814                        & -2.817273                       & 4.102827                        & -3.176874                       & 1.008829                        & 0.247031                        & {\color[HTML]{00B0F0} 0.0000004}       \\
\hline\hline
\end{tabular}}\\
The exponents of the corresponding (Gaussian) polarization functions are shown in green.
\end{table}
}

{\scriptsize
\begin{table}[H]
\caption{Conventional ANO contractions of the 7ZaPa primitive set for carbon}
\resizebox{\textwidth}{!}{\begin{tabular}{|l|l|rrrrrrr||l|}
\hline\hline

d & C   & {\color[HTML]{00B050} 5.67E+00} & {\color[HTML]{00B050} 2.83E+00} & {\color[HTML]{00B050} 1.41E+00} & {\color[HTML]{00B050} 7.05E-01} & {\color[HTML]{00B050} 3.52E-01} & {\color[HTML]{00B050} 1.75E-01} & {\color[HTML]{00B050} 8.76E-02} & NO occ                                 \\
c & 1.7 & 0.009138                        & 0.022196                        & 0.144379                        & 0.361068                        & 0.409154                        & 0.19832                         & 0.038143                        & {\color[HTML]{00B0F0} 0.0050464}       \\
c & 1.7 & 0.011602                        & 0.027832                        & 0.607837                        & 0.465256                        & -0.543455                       & -0.401857                       & -0.083106                       & {\color[HTML]{00B0F0} 0.000102}        \\
c & 1.7 & 0.009266                        & 0.39913                         & 0.94977                         & -1.378619                       & -0.125486                       & 0.649997                        & 0.170355                        & {\color[HTML]{00B0F0} 0.0000143}       \\
c & 1.7 & 0.079713                        & 1.390553                        & -1.41344                        & -0.498644                       & 1.835338                        & -1.093839                       & -0.147692                       & {\color[HTML]{00B0F0} 0.0000022}       \\
c & 1.7 & 0.819814                        & 0.344919                        & -2.726979                       & 4.019961                        & -3.274642                       & 1.20028                         & 0.176604                        & {\color[HTML]{00B0F0} 0.0000005}       \\
\hline\hline
\end{tabular}}\\
The exponents of the corresponding (Gaussian) polarization functions are shown in green.
\end{table}
}

{\scriptsize
\begin{table}[H]
\caption{Conventional ANO contractions of the 7ZaPa primitive set for nitrogen}
\resizebox{\textwidth}{!}{\begin{tabular}{|l|l|rrrrrrr||l|}
\hline\hline

d & N   & {\color[HTML]{00B050} 6.85E+00} & {\color[HTML]{00B050} 3.44E+00} & {\color[HTML]{00B050} 1.72E+00} & {\color[HTML]{00B050} 8.64E-01} & {\color[HTML]{00B050} 4.33E-01} & {\color[HTML]{00B050} 2.17E-01} & {\color[HTML]{00B050} 1.09E-01} & NO occ                           \\
c & 1.7 & 0.015121                        & 0.038131                        & 0.203379                        & 0.394576                        & 0.36647                         & 0.14555                         & 0.024074                        & {\color[HTML]{00B0F0} 0.0048794} \\
c & 1.7 & 0.023102                        & 0.131785                        & 0.699508                        & 0.181656                        & -0.58457                        & -0.31403                        & -0.052224                       & {\color[HTML]{00B0F0} 0.0001422} \\
c & 1.7 & 0.020377                        & 0.685622                        & 0.486743                        & -1.495451                       & 0.336902                        & 0.516639                        & 0.122947                        & {\color[HTML]{00B0F0} 0.0000171} \\
c & 1.7 & 0.2811                          & 1.315323                        & -2.197929                       & 0.76765                         & 1.015134                        & -0.939329                       & -0.110955                       & {\color[HTML]{00B0F0} 0.0000029} \\
c & 1.7 & 1.346608                        & -1.137178                       & -0.995173                       & 2.984038                        & -3.181757                       & 1.518395                        & 0.034767                        & {\color[HTML]{00B0F0} 0.0000006} \\
\hline\hline
\end{tabular}}\\
The exponents of the corresponding (Gaussian) polarization functions are shown in green.
\end{table}
}

{\scriptsize
\begin{table}[H]
\caption{Conventional ANO contractions of the 7ZaPa primitive set for oxygen}
\resizebox{\textwidth}{!}{\begin{tabular}{|l|l|rrrrrrr||l|}
\hline\hline

d & O   & {\color[HTML]{00B050} 9.46E+00} & {\color[HTML]{00B050} 4.75E+00} & {\color[HTML]{00B050} 2.39E+00} & {\color[HTML]{00B050} 1.20E+00} & {\color[HTML]{00B050} 6.02E-01} & {\color[HTML]{00B050} 3.02E-01} & {\color[HTML]{00B050} 1.52E-01} & NO occ                           \\
c & 1.7 & 0.015961                        & 0.037833                        & 0.202365                        & 0.395033                        & 0.364479                        & 0.145011                        & 0.02894                         & {\color[HTML]{00B0F0} 0.0045016} \\
c & 1.7 & 0.025608                        & 0.128218                        & 0.621125                        & 0.25521                         & -0.531671                       & -0.372347                       & -0.090942                       & {\color[HTML]{00B0F0} 0.0002604} \\
c & 1.7 & 0.042766                        & 0.556895                        & 0.628504                        & -1.357873                       & 0.062422                        & 0.565022                        & 0.204312                        & {\color[HTML]{00B0F0} 0.0000302} \\
c & 1.7 & 0.159851                        & 1.329349                        & -1.728451                       & -0.047795                       & 1.546331                        & -0.931421                       & -0.237271                       & {\color[HTML]{00B0F0} 0.0000049} \\
c & 1.7 & 0.97588                         & -0.110539                       & -2.24726                        & 3.690546                        & -2.997169                       & 0.915108                        & 0.339281                        & {\color[HTML]{00B0F0} 0.0000012} \\
\hline\hline
\end{tabular}}\\
The exponents of the corresponding (Gaussian) polarization functions are shown in green.
\end{table}
}

{\scriptsize
\begin{table}[H]
\caption{Conventional ANO contractions of the 7ZaPa primitive set for fluorine}
\resizebox{\textwidth}{!}{\begin{tabular}{|l|l|rrrrrrr||l|}
\hline\hline

d & F   & {\color[HTML]{00B050} 1.64E+01} & {\color[HTML]{00B050} 7.68E+00} & {\color[HTML]{00B050} 3.61E+00} & {\color[HTML]{00B050} 1.70E+00} & {\color[HTML]{00B050} 7.96E-01} & {\color[HTML]{00B050} 3.74E-01} & {\color[HTML]{00B050} 1.76E-01} & NO occ                           \\
c & 1.7 & 0.009617                        & 0.028361                        & 0.171799                        & 0.422771                        & 0.396038                        & 0.14236                         & 0.020957                        & {\color[HTML]{00B0F0} 0.0041892} \\
c & 1.7 & 0.017308                        & 0.074781                        & 0.563509                        & 0.393627                        & -0.564414                       & -0.388567                       & -0.070687                       & {\color[HTML]{00B0F0} 0.0003352} \\
c & 1.7 & 0.01653                         & 0.348403                        & 0.897547                        & -1.266788                       & -0.091235                       & 0.671789                        & 0.126668                        & {\color[HTML]{00B0F0} 0.0000369} \\
c & 1.7 & 0.003385                        & 1.189449                        & -0.916308                       & -0.833209                       & 1.89549                         & -1.129186                       & -0.119743                       & {\color[HTML]{00B0F0} 0.0000061} \\
c & 1.7 & 0.340636                        & 1.158039                        & -2.997826                       & 3.572389                        & -2.678009                       & 0.938132                        & 0.206736                        & {\color[HTML]{00B0F0} 0.0000015} \\
\hline\hline
\end{tabular}}\\
The exponents of the corresponding (Gaussian) polarization functions are shown in green.
\end{table}
}

\begin{table}[H]
\caption{Conventional ANO contractions of the 7ZaPa primitive set for aluminum}
\centering
\resizebox{\textwidth}{!}{%
\begin{tabular}{|l|l|rrrrrrrr||l|}
\hline

d & AL  & {\color[HTML]{00B050} 5.53E+00} & {\color[HTML]{00B050} 1.77E+00} & {\color[HTML]{00B050} 9.33E-01} & {\color[HTML]{00B050} 4.91E-01} & {\color[HTML]{00B050} 2.58E-01} & {\color[HTML]{00B050} 1.36E-01} & {\color[HTML]{00B050} 7.13E-02} & {\color[HTML]{00B050} 3.75E-02} & NO occ                                 \\
c & 1.8 & 0.005601                        & 0.01418                         & 0.010585                        & 0.123371                        & 0.361084                        & 0.413343                        & 0.196407                        & 0.034357                        & {\color[HTML]{0F9ED5} 0.0074976}       \\
c & 1.8 & -0.006326                       & -0.0001                         & 0.102124                        & -0.539405                       & -0.644944                       & 0.455358                        & 0.542707                        & 0.134203                        & {\color[HTML]{0F9ED5} 0.0001221}       \\
c & 1.8 & -0.009878                       & -0.179816                       & -0.02447                        & 1.871093                        & -1.682719                       & -0.412664                       & 0.734764                        & 0.230587                        & {\color[HTML]{0F9ED5} 0.0000086}       \\
c & 1.8 & 0.07853                         & -0.601081                       & 2.054669                        & -0.813618                       & -1.998275                       & 2.875414                        & -1.276347                       & -0.194672                       & {\color[HTML]{0F9ED5} 0.0000012}       \\
c & 1.8 & 0.436777                        & 0.413995                        & 1.095531                        & -3.225861                       & 4.08456                         & -3.274317                       & 1.357962                        & -0.038447                       & {\color[HTML]{0F9ED5} 0.0000002}       \\
\hline\hline

\end{tabular}
}\\
The exponents of the corresponding (Gaussian) polarization functions are shown in green.
\end{table}

\begin{table}[H]
\caption{Conventional ANO contractions of the 7ZaPa primitive set for silicon}
\centering
\resizebox{\textwidth}{!}{%
\begin{tabular}{|l|l|rrrrrrrr||l|}
\hline

d & SI  & {\color[HTML]{00B050} 9.54E+00} & {\color[HTML]{00B050} 3.30E+00} & {\color[HTML]{00B050} 1.65E+00} & {\color[HTML]{00B050} 8.20E-01} & {\color[HTML]{00B050} 4.09E-01} & {\color[HTML]{00B050} 2.04E-01} & {\color[HTML]{00B050} 1.01E-01} & {\color[HTML]{00B050} 5.06E-02} & NO occ                                 \\
c & 1.8 & 0.004033                        & 0.01227                         & 0.012952                        & 0.089024                        & 0.365306                        & 0.45268                         & 0.198815                        & 0.025894                        & {\color[HTML]{00B0F0} 0.010879}        \\
c & 1.8 & 0.002766                        & -0.014691                       & 0.131794                        & -0.342143                       & -0.87823                        & 0.510108                        & 0.56161                         & 0.101937                        & {\color[HTML]{00B0F0} 0.0001601}       \\
c & 1.8 & -0.025905                       & -0.001987                       & -0.464222                       & 1.86245                         & -1.016995                       & -0.986                          & 1.010816                        & 0.128206                        & {\color[HTML]{00B0F0} 0.0000119}       \\
c & 1.8 & 0.028097                        & -0.269218                       & 0.510882                        & 1.534257                        & -3.607039                       & 3.459708                        & -1.479097                       & -0.124889                       & {\color[HTML]{00B0F0} 0.0000017}       \\
c & 1.8 & 0.391636                        & 0.552618                        & 0.715109                        & -1.814022                       & 1.957726                        & -1.505572                       & 0.633276                        & -0.025648                       & {\color[HTML]{00B0F0} 0.0000004}       \\
\hline\hline

\end{tabular}
}\\
The exponents of the corresponding (Gaussian) polarization functions are shown in green.
\end{table}

\begin{table}[H]
\caption{Conventional ANO contractions of the 7ZaPa primitive set for phosphorus}
\centering
\resizebox{\textwidth}{!}{%
\begin{tabular}{|l|l|rrrrrrrr||l|}
\hline

d & P   & {\color[HTML]{00B050} 1.24E+01} & {\color[HTML]{00B050} 4.69E+00} & {\color[HTML]{00B050} 2.30E+00} & {\color[HTML]{00B050} 1.13E+00} & {\color[HTML]{00B050} 5.54E-01} & {\color[HTML]{00B050} 2.72E-01} & {\color[HTML]{00B050} 1.33E-01} & {\color[HTML]{00B050} 6.54E-02} & NO occ                                 \\
c & 1.8 & 0.00399                         & 0.011218                        & 0.016054                        & 0.087494                        & 0.378759                        & 0.457439                        & 0.185415                        & 0.021475                        & {\color[HTML]{00B0F0} 0.0123492}       \\
c & 1.8 & 0.007161                        & -0.014312                       & 0.138221                        & -0.311161                       & -0.934671                       & 0.586418                        & 0.536221                        & 0.0823                          & {\color[HTML]{00B0F0} 0.0001959}       \\
c & 1.8 & -0.032882                       & 0.035548                        & -0.501972                       & 1.808889                        & -0.922321                       & -1.045759                       & 1.082808                        & 0.08194                         & {\color[HTML]{00B0F0} 0.0000151}       \\
c & 1.8 & 0.010401                        & 0.257404                        & -0.208031                       & -1.937507                       & 3.830543                        & -3.58749                        & 1.612915                        & 0.064834                        & {\color[HTML]{00B0F0} 0.000002}        \\
c & 1.8 & 0.427727                        & 0.41083                         & 0.825893                        & -1.495479                       & 1.267837                        & -0.778823                       & 0.170138                        & 0.103226                        & {\color[HTML]{00B0F0} 0.0000006}       \\
\hline\hline

\end{tabular}
}\\
The exponents of the corresponding (Gaussian) polarization functions are shown in green.
\end{table}

\begin{table}[H]
\caption{Conventional ANO contractions of the 7ZaPa primitive set for sulfur}
\centering
\resizebox{\textwidth}{!}{%
\begin{tabular}{|l|l|rrrrrrrr||l|}
\hline

d & S   & {\color[HTML]{00B050} 1.51E+01} & {\color[HTML]{00B050} 5.76E+00} & {\color[HTML]{00B050} 2.81E+00} & {\color[HTML]{00B050} 1.37E+00} & {\color[HTML]{00B050} 6.71E-01} & {\color[HTML]{00B050} 3.28E-01} & {\color[HTML]{00B050} 1.60E-01} & {\color[HTML]{00B050} 7.81E-02} & NO occ                                 \\
c & 1.8 & 0.004546                        & 0.012616                        & 0.019457                        & 0.099163                        & 0.399836                        & 0.446641                        & 0.164581                        & 0.018531                        & {\color[HTML]{00B0F0} 0.014454}        \\
c & 1.8 & 0.006299                        & -0.014788                       & 0.111847                        & -0.338899                       & -0.874538                       & 0.608827                        & 0.516477                        & 0.082644                        & {\color[HTML]{00B0F0} 0.0003913}       \\
c & 1.8 & -0.031299                       & 0.028406                        & -0.438523                       & 1.752096                        & -0.966723                       & -0.975124                       & 1.070705                        & 0.081412                        & {\color[HTML]{00B0F0} 0.0000286}       \\
c & 1.8 & -0.00811                        & 0.193904                        & -0.339489                       & -1.62275                        & 3.527065                        & -3.383215                       & 1.524449                        & 0.080881                        & {\color[HTML]{00B0F0} 0.0000043}       \\
c & 1.8 & 0.383845                        & 0.232338                        & 1.325101                        & -2.320867                       & 2.175138                        & -1.504222                       & 0.544815                        & 0.041181                        & {\color[HTML]{00B0F0} 0.0000011}       \\
\hline\hline

\end{tabular}
}\\
The exponents of the corresponding (Gaussian) polarization functions are shown in green.
\end{table}

\begin{table}[H]
\caption{Conventional ANO contractions of the 7ZaPa primitive set for chlorine}
\centering
\resizebox{\textwidth}{!}{%
\begin{tabular}{|l|l|rrrrrrrr||l|}
\hline

d & CL  & {\color[HTML]{00B050} 2.40E+01} & {\color[HTML]{00B050} 7.68E+00} & {\color[HTML]{00B050} 3.69E+00} & {\color[HTML]{00B050} 1.78E+00} & {\color[HTML]{00B050} 8.54E-01} & {\color[HTML]{00B050} 4.11E-01} & {\color[HTML]{00B050} 1.98E-01} & {\color[HTML]{00B050} 9.50E-02} & NO occ                                 \\
c & 1.8 & 0.003484                        & 0.0141                          & 0.019466                        & 0.093925                        & 0.397613                        & 0.457812                        & 0.163527                        & 0.018398                        & {\color[HTML]{00B0F0} 0.0159926}       \\
c & 1.8 & 0.004571                        & -0.01098                        & 0.094691                        & -0.279386                       & -0.920074                       & 0.61129                         & 0.523576                        & 0.074933                        & {\color[HTML]{00B0F0} 0.000513}        \\
c & 1.8 & -0.022424                       & 0.033646                        & -0.429425                       & 1.619185                        & -0.740991                       & -1.131411                       & 1.146555                        & 0.05079                         & {\color[HTML]{00B0F0} 0.0000384}       \\
c & 1.8 & -0.002734                       & 0.106951                        & -0.139305                       & -1.814419                       & 3.525178                        & -3.271565                       & 1.482101                        & 0.064133                        & {\color[HTML]{00B0F0} 0.0000058}       \\
c & 1.8 & 0.301259                        & 0.441888                        & 1.043965                        & -1.957362                       & 1.804547                        & -1.208921                       & 0.37685                         & 0.095191                        & {\color[HTML]{00B0F0} 0.0000016}       \\
\hline\hline

\end{tabular}
}\\
The exponents of the corresponding (Gaussian) polarization functions are shown in green.
\end{table}

\newpage
\section{Pseudo-ANO contractions of the 7ZaPa primitive set}
 The 7ZaPa primitive exponents (Ranasinghe and Petersson, J. Chem. Phys. \textbf{138}, 144104 (2013)) are shown rounded to three significant figures; full-precision values are available from the Basis Set Exchange.

\begin{table}[H]
\caption{Pseudo-ANO contractions of the 7ZaPa primitive set for boron}
\resizebox{\textwidth}{!}{\begin{tabular}{|l|l|lllllll|}
\hline\hline

d & B   & {\color[HTML]{00B050} 3.64E+00} & {\color[HTML]{00B050} 1.84E+00} & {\color[HTML]{00B050} 9.33E-01} & {\color[HTML]{00B050} 4.72E-01} & {\color[HTML]{00B050} 2.39E-01} & {\color[HTML]{00B050} 1.21E-01} & {\color[HTML]{00B050} 6.13E-02} \\
c & 1.7 & 0.018706                        & 0.040894                        & 0.299307                        & 0.815042                        & 1                               & 0.487709                        & 0.086542                        \\
c & 1.7 & 0.015496                        & 0.002961                        & 0.862008                        & 1                               & 0                               & -0.038226                       & -0.017926                       \\
c & 1.7 & 0.022711                        & 0.177931                        & 1                               & 0                               & 0                               & 0.192467                        & 0.03671                         \\
c & 1.7 & 0.202431                        & 1                               & 0                               & 0                               & 0                               & -0.347239                       & 0.005409                        \\
c & 1.7 & 0.255162                        & 0                               & 0                               & 0                               & 0                               & 1                               & 0.109267                        \\
c & 1.7 & 1                               & 0                               & 0                               & 0                               & 0                               & 0                               & -0.026626                       \\
\hline\hline
\end{tabular}}\\
The exponents of the corresponding (Gaussian) polarization functions are shown in green.
\end{table}

\begin{table}[H]
\caption{Pseudo-ANO contractions of the 7ZaPa primitive set for carbon}
\resizebox{\textwidth}{!}{\begin{tabular}{|l|l|lllllll|}
\hline\hline
d & C   & {\color[HTML]{00B050} 5.67E+00} & {\color[HTML]{00B050} 2.83E+00} & {\color[HTML]{00B050} 1.41E+00} & {\color[HTML]{00B050} 7.05E-01} & {\color[HTML]{00B050} 3.52E-01} & {\color[HTML]{00B050} 1.75E-01} & {\color[HTML]{00B050} 8.76E-02} \\
c & 1.7 & 0.023115                        & 0.057656                        & 0.36395                         & 0.898041                        & 1                               & 0.433135                        & 0.067542                        \\
c & 1.7 & 0.037452                        & 0.104767                        & 1                               & 0.951957                        & 0                               & -0.03312                        & -0.010965                       \\
c & 1.7 & 0.00214                         & -0.181908                       & 0                               & 1                               & 0                               & -0.098891                       & -0.027448                       \\
c & 1.7 & 0.177514                        & 1                               & 0                               & 0                               & 0                               & -0.470512                       & -0.006362                       \\
c & 1.7 & 0.25065                         & 0                               & 0                               & 0                               & 0                               & 1                               & 0.058607                        \\
c & 1.7 & 1                               & 0                               & 0                               & 0                               & 0                               & 0                               & -0.063555                       \\
\hline\hline
\end{tabular}}\\
The exponents of the corresponding (Gaussian) polarization functions are shown in green.
\end{table}

\begin{table}[H]
\caption{Pseudo-ANO contractions of the 7ZaPa primitive set for nitrogen}
\resizebox{\textwidth}{!}{\begin{tabular}{|l|l|lllllll|}
\hline\hline

d & N   & {\color[HTML]{00B050} 6.85E+00} & {\color[HTML]{00B050} 3.44E+00} & {\color[HTML]{00B050} 1.72E+00} & {\color[HTML]{00B050} 8.64E-01} & {\color[HTML]{00B050} 4.33E-01} & {\color[HTML]{00B050} 2.17E-01} & {\color[HTML]{00B050} 1.09E-01} \\
c & 1.7 & 0.040035                        & 0.103462                        & 0.528689                        & 1                               & 0.884104                        & 0.294345                        & 0.039459                        \\
c & 1.7 & 0.046329                        & 0.241955                        & 1                               & 0                               & -0.808965                       & -0.3026                         & -0.041047                       \\
c & 1.7 & 0.030727                        & 0.300563                        & 0                               & 0                               & 1                               & 0.330389                        & 0.061212                        \\
c & 1.7 & 0.261713                        & 1                               & 0                               & 0                               & 0                               & -0.210873                       & 0.006982                        \\
c & 1.7 & 1                               & 0                               & 0                               & 0                               & 0                               & 0.750103                        & -0.013896                       \\
c & 1.7 & 0                               & 0                               & 0                               & 0                               & 0                               & 1                               & 0.059124                        \\
\hline\hline
\end{tabular}}\\
The exponents of the corresponding (Gaussian) polarization functions are shown in green.
\end{table}

\begin{table}[H]
\caption{Pseudo-ANO contractions of the 7ZaPa primitive set for oxygen}
\resizebox{\textwidth}{!}{\begin{tabular}{|l|l|lllllll|}
\hline\hline

d & O   & {\color[HTML]{00B050} 9.46E+00} & {\color[HTML]{00B050} 4.75E+00} & {\color[HTML]{00B050} 2.39E+00} & {\color[HTML]{00B050} 1.20E+00} & {\color[HTML]{00B050} 6.02E-01} & {\color[HTML]{00B050} 3.02E-01} & {\color[HTML]{00B050} 1.52E-01} \\
c & 1.7 & 0.043878                        & 0.110195                        & 0.552617                        & 1                               & 0.821917                        & 0.258663                        & 0.044602                        \\
c & 1.7 & -0.049754                       & -0.245794                       & -0.971254                       & 0                               & 1                               & 0.419605                        & 0.071373                        \\
c & 1.7 & 0.079055                        & 0.453838                        & 1                               & 0                               & 0                               & -0.016783                       & 0.008485                        \\
c & 1.7 & 0.235569                        & 1                               & 0                               & 0                               & 0                               & -0.215827                       & -0.002057                       \\
c & 1.7 & 0.208289                        & 0                               & 0                               & 0                               & 0                               & 1                               & 0.088506                        \\
c & 1.7 & 1                               & 0                               & 0                               & 0                               & 0                               & 0                               & -0.028652                       \\
\hline\hline
\end{tabular}}\\
The exponents of the corresponding (Gaussian) polarization functions are shown in green.
\end{table}

\begin{table}[H]
\caption{Pseudo-ANO contractions of the 7ZaPa primitive set for fluorine}
\resizebox{\textwidth}{!}{\begin{tabular}{|l|l|lllllll|}
\hline\hline

d & F   & {\color[HTML]{00B050} 1.64E+01} & {\color[HTML]{00B050} 7.68E+00} & {\color[HTML]{00B050} 3.61E+00} & {\color[HTML]{00B050} 1.70E+00} & {\color[HTML]{00B050} 7.96E-01} & {\color[HTML]{00B050} 3.74E-01} & {\color[HTML]{00B050} 1.76E-01} \\
c & 1.7 & 0.02513                         & 0.080154                        & 0.461415                        & 1                               & 0.770123                        & 0.205584                        & 0.028468                        \\
c & 1.7 & -0.0197                         & -0.103107                       & -0.648196                       & 0                               & 1                               & 0.353363                        & 0.052012                        \\
c & 1.7 & 0.038644                        & 0.323531                        & 1                               & 0                               & 0                               & 0.04387                         & 0.003741                        \\
c & 1.7 & 0.079238                        & 1                               & 0                               & 0                               & 0                               & -0.374464                       & 0.012707                        \\
c & 1.7 & 0.04498                         & 0                               & 0                               & 0                               & 0                               & 1                               & 0.064449                        \\
c & 1.7 & 1                               & 0                               & 0                               & 0                               & 0                               & 0                               & 0.004209                        \\
\hline\hline
\end{tabular}}\\
The exponents of the corresponding (Gaussian) polarization functions are shown in green.
\end{table}

\begin{table}[H]
\caption{Pseudo-ANO contractions of the 7ZaPa primitive set for aluminum}
\resizebox{\textwidth}{!}{\begin{tabular}{|l|l|llllllll|}
\hline\hline

d & AL  & {\color[HTML]{00B050} 5.53E+00} & {\color[HTML]{00B050} 1.77E+00} & {\color[HTML]{00B050} 9.33E-01} & {\color[HTML]{00B050} 4.91E-01} & {\color[HTML]{00B050} 2.58E-01} & {\color[HTML]{00B050} 1.36E-01} & {\color[HTML]{00B050} 7.13E-02} & {\color[HTML]{00B050} 3.75E-02} \\
c & 1.8 & 0.014075                        & 0.035977                        & 0.026847                        & 0.302861                        & 0.895109                        & 1                               & 0.431708                        & 0.067941                        \\
c & 1.8 & 0.011602                        & 0.020657                        & -0.085259                       & 0.672713                        & 1                               & 0                               & -0.166377                       & -0.042576                       \\
c & 1.8 & 0.001953                        & -0.042169                       & -0.065808                       & 1                               & 0                               & 0                               & 0.023828                        & 0.007926                        \\
c & 1.8 & -0.021031                       & 0.260894                        & -0.693046                       & 0                               & 0                               & 0                               & 1                               & 0.07851                         \\
c & 1.8 & -0.138265                       & -0.471544                       & 1                               & 0                               & 0                               & 0                               & 0                               & 0.029195                        \\
\hline\hline
\end{tabular}}\\
The exponents of the corresponding (Gaussian) polarization functions are shown in green.
\end{table}

\begin{table}[H]
\caption{Pseudo-ANO contractions of the 7ZaPa primitive set for silicon}
\resizebox{\textwidth}{!}{\begin{tabular}{|l|l|llllllll|}
\hline\hline

d & SI  & {\color[HTML]{00B050} 9.54E+00} & {\color[HTML]{00B050} 3.30E+00} & {\color[HTML]{00B050} 1.65E+00} & {\color[HTML]{00B050} 8.20E-01} & {\color[HTML]{00B050} 4.09E-01} & {\color[HTML]{00B050} 2.04E-01} & {\color[HTML]{00B050} 1.01E-01} & {\color[HTML]{00B050} 5.06E-02} \\
c & 1.8 & 0.009161                        & 0.027953                        & 0.030682                        & 0.200791                        & 0.818119                        & 1                               & 0.407679                        & 0.048408                        \\
c & 1.8 & 0.001259                        & 0.022803                        & -0.080375                       & 0.359389                        & 1                               & 0                               & -0.139762                       & -0.027921                       \\
c & 1.8 & -0.013242                       & 0.009927                        & -0.233538                       & 1                               & 0                               & 0                               & 0.316309                        & 0.007425                        \\
c & 1.8 & 0.004158                        & 0.072442                        & -0.122367                       & 0                               & 0                               & 0                               & 1                               & 0.04268                         \\
c & 1.8 & -0.389727                       & -0.889432                       & 1                               & 0                               & 0                               & 0                               & 0                               & 0.07331                         \\
\hline\hline
\end{tabular}}\\
The exponents of the corresponding (Gaussian) polarization functions are shown in green.
\end{table}

\begin{table}[H]
\caption{Pseudo-ANO contractions of the 7ZaPa primitive set for phosphorus}
\resizebox{\textwidth}{!}{\begin{tabular}{|l|l|llllllll|}
\hline\hline

d & P   & {\color[HTML]{00B050} 1.24E+01} & {\color[HTML]{00B050} 4.69E+00} & {\color[HTML]{00B050} 2.30E+00} & {\color[HTML]{00B050} 1.13E+00} & {\color[HTML]{00B050} 5.54E-01} & {\color[HTML]{00B050} 2.72E-01} & {\color[HTML]{00B050} 1.33E-01} & {\color[HTML]{00B050} 6.54E-02} \\
c & 1.8 & 0.00895                         & 0.025228                        & 0.037727                        & 0.19906                         & 0.840291                        & 1                               & 0.378635                        & 0.040052                        \\
c & 1.8 & -0.001271                       & 0.020409                        & -0.071248                       & 0.319385                        & 1                               & 0                               & -0.112393                       & -0.020373                       \\
c & 1.8 & -0.018593                       & 0.021989                        & -0.253852                       & 1                               & 0                               & 0                               & 0.373375                        & -0.002841                       \\
c & 1.8 & 0.016032                        & 0.090346                        & -0.100755                       & 0                               & 0                               & 0                               & 1                               & 0.011931                        \\
c & 1.8 & 0.536544                        & 1                               & -0.9736                         & 0                               & 0                               & 0                               & 0                               & -0.051745                       \\
\hline\hline
\end{tabular}}\\
The exponents of the corresponding (Gaussian) polarization functions are shown in green.
\end{table}

\begin{table}[H]
\caption{Pseudo-ANO contractions of the 7ZaPa primitive set for sulfur}
\resizebox{\textwidth}{!}{\begin{tabular}{|l|l|llllllll|}
\hline\hline

d & S   & {\color[HTML]{00B050} 1.51E+01} & {\color[HTML]{00B050} 5.76E+00} & {\color[HTML]{00B050} 2.81E+00} & {\color[HTML]{00B050} 1.37E+00} & {\color[HTML]{00B050} 6.71E-01} & {\color[HTML]{00B050} 3.28E-01} & {\color[HTML]{00B050} 1.60E-01} & {\color[HTML]{00B050} 7.81E-02} \\
c & 1.8 & 0.010564                        & 0.029402                        & 0.046855                        & 0.238555                        & 0.92366                         & 1                               & 0.338103                        & 0.035607                        \\
c & 1.8 & 0.000791                        & 0.023485                        & -0.046477                       & 0.353484                        & 1                               & 0                               & -0.129754                       & -0.022221                       \\
c & 1.8 & -0.016202                       & 0.023524                        & -0.216343                       & 1                               & 0                               & 0                               & 0.389102                        & -0.005053                       \\
c & 1.8 & 0.00551                         & 0.056272                        & -0.095875                       & 0                               & 0                               & 0                               & 1                               & 0.021973                        \\
c & 1.8 & 0.685919                        & 1                               & -0.978301                       & 0                               & 0                               & 0                               & 0                               & -0.056132                       \\
\hline\hline
\end{tabular}}\\
The exponents of the corresponding (Gaussian) polarization functions are shown in green.
\end{table}

\begin{table}[H]
\caption{Pseudo-ANO contractions of the 7ZaPa primitive set for chlorine}
\resizebox{\textwidth}{!}{\begin{tabular}{|l|l|llllllll|}
\hline\hline

d & CL  & {\color[HTML]{00B050} 2.40E+01} & {\color[HTML]{00B050} 7.68E+00} & {\color[HTML]{00B050} 3.69E+00} & {\color[HTML]{00B050} 1.78E+00} & {\color[HTML]{00B050} 8.54E-01} & {\color[HTML]{00B050} 4.11E-01} & {\color[HTML]{00B050} 1.98E-01} & {\color[HTML]{00B050} 9.50E-02} \\
c & 1.8 & 0.007935                        & 0.032196                        & 0.045523                        & 0.224813                        & 0.900454                        & 1                               & 0.327524                        & 0.034864                        \\
c & 1.8 & 0.000749                        & 0.021747                        & -0.03615                        & 0.304769                        & 1                               & 0                               & -0.134992                       & -0.018653                       \\
c & 1.8 & -0.01279                        & 0.023579                        & -0.224278                       & 1                               & 0                               & 0                               & 0.520146                        & -0.012545                       \\
c & 1.8 & 0.004864                        & 0.046084                        & -0.064702                       & 0                               & 0                               & 0                               & 1                               & 0.011382                        \\
c & 1.8 & 0.442774                        & 1                               & -0.699556                       & 0                               & 0                               & 0                               & 0                               & -0.029935                       \\
\hline\hline
\end{tabular}}\\
The exponents of the corresponding (Gaussian) polarization functions are shown in green.
\end{table}

\section{Alternative pANO scheme for hydrogen from cc-pV5Z-F12}

{\scriptsize
\begin{table}[H]
\caption{RMSD (\kcalmol) from df-MP2-F12/cc-pV5Z-F12 for the 126 hydrogen-containing species of the W4-17 thermochemical benchmark as a function of the hydrogen pANO-F12 contraction scheme ($\gamma=1.2$), with contractions derived from cc-pV5Z-F12 rather than cc-pV5Z-F12rev2. This scheme was superseded by the cc-pV5Z-F12rev2-based one in the main text (Table 4). s$_i$ denotes the $s$ contractions, their number being given by the column header.}
\label{tab:RMSD-Hygrogen-V5Z-F12}
\begin{tabular}{|l|l|l|l|l|l|l|l|}
\hline\hline
Contractions & 2s & 3s & 4s & 5s & 6s & 7s & 8s \\
\hline
$s_i$  & 0.393 & 0.370 & 0.312 & 0.307 & 0.294 & 0.292 & 0.283 \\

$s_i$ + 1p(pANO) & 0.131 & 0.123 & 0.113 & 0.109 & 0.100 & 0.098 & 0.093 \\

$s_i$ + 2p(pANO) &  & 0.094 & 0.090 & 0.088 & 0.085 & 0.084 & 0.077 \\
$s_i$ + 2p+1d(pANO) &  & 0.067 & 0.064 & 0.062 & 0.060 & 0.059 & 0.056 \\

$s_i$ + 3p(pANO) &  &  & 0.085 & 0.083 & 0.080 & 0.077 & 0.069 \\
$s_i$ + 3p+1d(pANO) &  &  & 0.058 & 0.056 & 0.054 & 0.050 & 0.048 \\
$s_i$ + 3p+2d(pANO) &  &  & 0.041 & 0.041 & 0.039 & 0.037 & 0.035 \\
$s_i$ + 3p+2d+1f(pANO) &  &  & 0.033 & 0.033 & 0.032 & 0.030 & 0.028 \\

$s_i$ + 4p(pANO) &  &  &  & 0.075 & 0.071 & 0.069 & 0.062 \\
$s_i$ + 4p+1d(pANO) &  &  &  & 0.048 & 0.047 & 0.043 & 0.041 \\
$s_i$ + 4p+2d(pANO) &  &  &  & 0.037 & 0.036 & 0.034 & 0.032 \\
$s_i$ + 4p+2d+1f(pANO) &  &  &  & 0.030 & 0.029 & 0.027 & 0.025 \\
$s_i$ + 4p+3d(pANO) &  &  &  & 0.033 & 0.031 & 0.029 & 0.028 \\
$s_i$ + 4p+3d+1f(pANO) &  &  &  & 0.025 & 0.024 & 0.022 & 0.021 \\
$s_i$ + 4p+3d+2f(pANO) &  &  &  & 0.017 & 0.016 & 0.015 & 0.014 \\
$s_i$ + 4p+3d+2f+1g(pANO) &  &  &  & 0.013 & 0.012 & 0.011 & 0.010 \\
\hline\hline
5Z-F12 on non H; cc-pVDZ-F12 & \multicolumn{7}{c|}{0.176} \\
5Z-F12 on non H; cc-pVTZ-F12 & \multicolumn{7}{c|}{0.065} \\
5Z-F12 on non H; cc-pVQZ-F12 & \multicolumn{7}{c|}{0.018} \\
cc-pV5Z-F12 & \multicolumn{7}{c|}{REF} \\
\hline
\end{tabular}

\end{table}
}